\documentclass[fleqn,usenatbib,useAMS]{mnras}

\usepackage{newtxtext,newtxmath}

\usepackage[T1]{fontenc}

\DeclareRobustCommand{\VAN}[3]{#2}
\let\VANthebibliography\thebibliography
\def\thebibliography{\DeclareRobustCommand{\VAN}[3]{##3}\VANthebibliography}

\usepackage{hyperref}
\usepackage{subcaption}
\usepackage{graphicx}	
\usepackage{amsmath}	
\usepackage{tablefootnote}
\usepackage{longtable}
\usepackage{lscape}

\usepackage[pagewise]{lineno}

\def\farcs{$^{\prime\prime}$}
\def\alerce{ALeRCE}
\def\rein{$\theta_{\rm{Ein}}$}

\def\phs{\phantom{$-$}}   
 
\def\phn{\phantom{0}}

\def\h0{H$_{0}$}

\title[Lensed SNe in ZTF]{A search for gravitationally lensed supernovae within the Zwicky Transient Facility public survey}

\author[M. R. Magee et al.]{
M. R. Magee$^{1,2}$\thanks{E-mail: mrmagee.astro@gmail.com}, A. Sainz de Murieta$^{1}$, T. E. Collett$^{1}$, W. Enzi$^{1}$
\\
$^{1}$Institute of Cosmology and Gravitation, University of Portsmouth, Burnaby Road, Portsmouth, PO1 3FX, UK \\ 
$^{2}$Department of Physics, University of Warwick, Gibbet Hill Road, Coventry CV4 7AL, UK \\ 
}

\date{}

\pubyear{2023}

\begin{document}
\label{firstpage}
\pagerange{\pageref{firstpage}--\pageref{lastpage}}
\maketitle

\begin{abstract}
Strong gravitational lensing of supernovae is exceedingly rare. To date, only a handful of lensed supernovae are known. Despite this, lensed supernovae have emerged as a promising method for measuring the current expansion rate of the Universe and breaking the Hubble tension. We present an extensive search for gravitationally lensed supernovae within the Zwicky Transient Facility (ZTF) public survey, covering 15\,215 transients with good light curves discovered during four years of observations. We crossmatch a catalogue of known and candidate lens galaxies with our sample and find three coincident sources, which were due to chance alignment. To search for supernovae magnified by unknown lenses, we test multiple methods suggested in the literature for the first time on real data. This includes selecting objects with extremely red colours, those that appear inconsistent with the host galaxy redshift, and those with bright absolute magnitudes inferred from the host galaxy redshift. We find a few hundred candidates, most of which are due to contamination from activate galactic nuclei, bogus detections, or unlensed supernovae. The false positive rate from these methods presents significant challenges for future surveys. In total, 132 unique transients were identified across all of our selection methods that required detailed manual rejection, which would be infeasible for larger samples. Overall, we do not find any compelling candidates for lensed supernovae, which is broadly consistent with previous estimates for the rate of lensed supernovae from the ZTF public survey alone and the number expected to pass the selection cuts we apply.
\end{abstract}




\section{Introduction}
\label{sect:intro}

Strong gravitational lensing occurs when a sufficiently massive foreground galaxy or cluster distorts space-time to such a degree that multiple images of any well-aligned background source can be formed (see e.g. \citealt{treu--10}). Depending on the gravitational potential of the galaxy and overall geometry of the lens-source system these images can appear highly magnified, offering a unique look at high redshift objects in the Universe that may be otherwise unobservable. In addition, time delays between the lensed images, arising from the different paths travelled through the Universe, are sensitive to cosmological distances and hence the rate of expansion or \h0 \citep{refsdal--64}. Time delays are independent of both the cosmic microwave background (CMB; \citealt{planck18--20}) and the local distance ladder \citep{riess--22}, and have therefore been identified as a promising tool for confronting the 5$\sigma$ difference between \h0 measured via these methods.

\par

Measurements of time delays require that the lensed sources are time-varying and observed over sufficiently long timescales (i.e. longer than the time delay itself; \citealt{treu--16}). To date, most time delay measurements have been made using gravitationally lensed active galactic nuclei (AGN; e.g. \citealt{kochanek--06, eulaers--13, bonvin--17, million--20, wong--20}). While significant progress has been made in this area, the stochastic nature of AGN light curves and long observing times required present challenges for precision measurements of \h0. Alternatively, supernovae (SNe) are transient phenomena much more suited to time delay measurements \citep{oguri--19}. The light curves of SNe are generally much simpler than AGN and vary on shorter timescales, requiring less observational overhead. In addition, light from the SN will eventually fade away after a few weeks or months, allowing for detailed follow up and reconstruction of the lensed host galaxy. This is very challenging for the hosts of lensed AGN however as lensed AGN typically outshine the host galaxy.

\par

Gravitationally lensed SNe (glSNe) offer many advantages over lensed AGN, but are currently much rarer. Whilst glSNe were first proposed as a tool to measure \h0 \citep{refsdal--64}, only a small handful (four) of SNe with multiple resolved images have been confirmed. \cite{kelly--15} report the discovery of `SN~Refsdal', a SN at $z = 1.49$ lensed by a massive galaxy cluster at $z = 0.54$ and producing a distinct Einstein cross configuration. Subsequent spectroscopic observations revealed that SN~Refsdal was likely a peculiar type II SN (SN~II; \citealt{kelly--16}). \cite{goobar--17} demonstrate that iPTF16geu was a multiply-imaged type Ia SN (SN~Ia) that occurred at $z = 0.41$ and was magnified in luminosity by a factor of $\gtrsim$50, although the initial discovery images from the intermediate Palomar Transient Facility (iPTF; \citealt{law--09}) were unresolved. \cite{rodney--21} present observations of `SN~Requiem', a glSN at $z = 1.95$ that they argue is consistent with being a SN~Ia. Most recently, \cite{goobar--22} present observations of `SN~Zwicky' a glSN~Ia at $z = 0.35$ discovered by the Zwicky Transient Facility (ZTF; \citealt{ztf, graham--19, masci--19, dekany--20}). A handful of other unresolved glSNe have also been proposed (seven), in addition to some targeted searches proving unsuccessful \citep{amanullah--11, quimby--13, patel--14, rodney--15, petrushevska--16, rubin--18, craig--21}. Aside from the difficulty in observing glSNe, the small sample size of objects may also be affected by our ability to correctly identify them. Even for those objects with spectroscopic observations, recognising glSNe from unresolved images can be challenging \citep{chornock--13, quimby--13, quimby--14}. Therefore the possibility remains that many more glSNe have been observed, but simply were not recognised as such.

\par

The simplest method of identifying glSNe is to monitor known lenses for new SNe (e.g. \citealt{craig--21}), however only a few thousand lens systems have been confirmed. \cite{quimby--14} propose selecting candidate glSNe from unresolved images based on their colours and magnitudes during the rising phase. As glSNe will be observed at higher redshifts than non-lensed SNe, the peak of the spectral energy distribution (SED) will be shifted out of bluer bands in the observer frame, producing redder observed colours. Hence, \cite{quimby--14} suggest searching for SNe that appear significantly redder than expected for a given apparent magnitude. Indeed, they show that the glSN~Ia PS1-10afx was redder around maximum light than non-lensed SNe~Ia by $\gtrsim$1.2~mag in $r - i$. \cite{goldstein--18--lens} discuss an alternative strategy for identifying glSNe.
Almost all SNe that occur in elliptical galaxies are SNe~Ia \citep{li--2011}. In addition, elliptical galaxies dominate the strong lensing cross-section in the Universe (\citealt{auger--09}). Therefore \cite{goldstein--18--lens} suggest that if a SN occurs near an elliptical galaxy, but is inconsistent with being a SN~Ia at the redshift of the elliptical, it could be a glSN at a higher redshift. Using this technique, \cite{goldstein--19} estimate the number of glSNe that can be discovered by ZTF and the upcoming Legacy Survey of Space and Time (LSST; \citealt{ivezic--19}), based on simulations of both surveys and populations of glSNe of all types. \cite{goldstein--19} estimate that the depth and coverage of LSST should allow $\gtrsim$340 glSNe to be discovered each year. For ZTF the observational depth is significantly shallower than LSST, however \cite{goldstein--19} estimate $\sim$1 -- 9 glSNe should be discovered per year, depending on which ZTF data is used for the search (whether this includes $i$-band, high-cadence, or only publicly available observations). The increased spatial resolution of LSST will also allow some of these glSNe to be resolved, which could further improve detection efficiency \citep{ramanah--22}.

\par

To date, only one glSN has been confirmed within ZTF, SN~Zwicky. As part of the Bright Transient Survey (BTS; \citealt{fremling--20}), ZTF aims to spectroscopically classify all SNe brighter than $m_{g,r} \sim 19$. SN~Zwicky became sufficiently bright to trigger automatic spectroscopic classification, which indicated that it was a SN~Ia at a redshift of $z = 0.35$. The light curve however, was clearly significantly brighter than expected for a SN~Ia at this redshift. Identification of glSNe may therefore also be possible by selecting SNe that appear to be significantly brighter than expected.

\par

Here we report on a systematic search for lensed SNe within the ZTF public survey using previously suggested identification methods. The construction of our initial sample is described in Sect.~\ref{sect:sample}. We first crossmatch our sample of transients against a list of confirmed or candidate lens systems in Sect.~\ref{sect:known}. In Sect.~\ref{sect:colour_method}, we apply the colour-based identification method presented by \cite{quimby--14} to our sample, while Sect.~\ref{sect:outlier_method} discusses the outlier-based method presented by \cite{goldstein--19}. Section~\ref{sect:luminosity_method} presents a luminosity-based selection method, following the discovery of SN~Zwicky. Finally, we discuss our results in Sect.~\ref{sect:discuss} and conclusions in Sect.~\ref{sect:conclusions}.

%

\section{The ZTF public sample}
\label{sect:sample}
Since beginning operations in 2018, ZTF observing time has been divided into a series of public and private surveys. The majority of the public time is dedicated to an all-sky survey, reaching typical limiting magnitudes of $\sim$20.5~mag in the $g$- and $r$-bands. During the first phase of operations, ZTF~I, the public survey operated with a typical cadence of 3\,d. Since the end of 2020, the public survey cadence has increased to 2\,d. Further details of the surveys conducted by ZTF are given by \cite{bellm--19}. 

\par

Our initial sample consists of all transients that were observed by ZTF and announced publicly on the Transient Name Server (TNS)\footnote{https://www.wis-tns.org/} from the beginning of the survey through to 2022 March 16, equalling approximately four years of survey operations. In total this resulted in 29\,242 transients, of which only 4\,820 were spectroscopically classified. We compared against the Million Quasars (Milliquas) Catalog v7.5\footnote{https://quasars.org/milliquas.htm} \citep{flesch--21} and the Open Cataclysmic Variable Catalog\footnote{https://depts.washington.edu/catvar/} \citep{guillochon--17} to remove all transients crossmatched within a 1.5\farcs{} radius, leaving 26\,575 transients.

\par

To ensure that our sample contains light curves of sufficient quality, such that they are scientifically useful and can be reliably fit with various models, we only selected objects for which there is sufficient data to perform template fitting. For each object in our sample, we queried the \alerce{}\footnote{https://alerce.readthedocs.io/en/latest/index.html} \citep{forster--21} database to retrieve the detections. We selected only those objects with detections in the $r$-band over at least four separate nights and spanning a period of at least one week. In addition, we also select objects for which the majority of difference image flux detections are positive. This reduced the number of transients to just over half, leaving 15\,215 objects in our sample. Given that glSNe are expected to be red, we do not set a requirement for any detections in the $g$-band.

\begin{table*}
\begin{center}
\caption{Summary of the cuts applied to the TNS public sample.}
\label{tab:cuts}
\resizebox{\textwidth}{!}{
\begin{tabular}{lrrr}
\hline
\textbf{Condition} & \textbf{No. of objects} & \textbf{Fraction} & \textbf{Total fraction} \\
& \textbf{remaining} & \textbf{removed} & \textbf{remaining} \\
\hline
\hline
Starting Sample                                                         & 29\,242 & -            & 100.000\%                \\
Not within 1.5\farcs{} of known galactic or stellar variable            & 26\,575 & \phn{}9.12\% & \phn{}90.880\%           \\
At least four nights of $r$-band detections over at least one week      & 15\,215 & 42.75\%      & \phn{}52.031\% \\
\hline
\multicolumn{4}{c}{Very red transients} \tabularnewline
\hline
Initial sample                                  & 15\,215               & 42.75\%   & \phn{}52.031\% \\
Red colours                                     & \phn{}\phn{}445       & 97.08\%   & \phn{}\phn{}1.522\% \\
Pre-maximum                                     & \phn{}\phn{}\phn{}84  & 81.12\%   & \phn{}\phn{}0.287\% \\
\hline
\multicolumn{4}{c}{Light curve inconsistent with SN~Ia at photometric redshift of nearby elliptical galaxy} \tabularnewline
\hline
Initial sample                                          & 15\,215                 & 42.75\%         & \phn{}52.031\% \\
Elliptical within 5\farcs{}                             & \phn{}1\,241            & 91.84\%         & \phn{}\phn{}4.244\% \\
Photo-$z$ for elliptical                                & \phn{}1\,203            & \phn{}3.06\%    & \phn{}\phn{}4.114\% \\
$\geq$1 5$\sigma$ outlier                               & \phn{}\phn{}310         & 74.23\%         & \phn{}\phn{}1.060\% \\
No clear AGN- or stellar-like variability               & \phn{}\phn{}133         & 57.10\%         & \phn{}\phn{}0.455\% \\
Outlier not due solely to non-detections                & \phn{}\phn{}\phn{}73    & 45.11\%         & \phn{}\phn{}0.250\% \\
Outlier not due to spurious detection                   & \phn{}\phn{}\phn{}56    & 23.29\%         & \phn{}\phn{}0.192\% \\
Outlier not due to phases beyond SALT2 template range   & \phn{}\phn{}\phn{}34    & 39.29\%         & \phn{}\phn{}0.116\% \\
No clear spiral arms in archival SDSS or PS1 imaging    & \phn{}\phn{}\phn{}29    & 14.71\%         & \phn{}\phn{}0.099\% \\
\hline
\multicolumn{4}{c}{Intrinsically luminous assuming host redshift} \tabularnewline
\hline
\multicolumn{4}{c}{PS1 photo-$z$ for host} \tabularnewline
\hline
Initial sample                                      & 15\,215                     & 42.75\%   & \phn{}52.031\% \\
Host within 5\farcs{}                               & 11\,056                     & 27.33\%   & \phn{}37.809\% \\
$\frac{\sigma_z}{1+z} \leq 0.05$                    & \phn{}7\,586                & 31.39\%   & \phn{}25.942\% \\
3$\sigma$ lower-bound $M_r \leq -21$                & \phn{}\phn{}\phn{}28        & 99.63\%   & \phn{}\phn{}0.096\% \\
No clear AGN- or stellar-like variability           & \phn{}\phn{}\phn{}19                          & 32.14\%   & \phn{}\phn{}0.065\% \\
\hline
\multicolumn{4}{c}{SDSS spec-$z$ for host} \tabularnewline
\hline
Initial sample                                      & 15\,215                     & 42.75\%   & \phn{}52.031\% \\
SDSS spec-$z$ host within 5\farcs{}                 & \phn{}1\,653                & 89.14\%   & \phn{}\phn{}5.653\% \\
3$\sigma$ lower-bound $M_r \leq -21$                & \phn{}\phn{}\phn{}\phn{}1         & 99.94\%   & \phn{}\phn{}0.003\% \\
\hline
\hline
\end{tabular}
}
\end{center}
\end{table*}

%

\section{Known lenses}
\label{sect:known}

We begin our search for lensed SNe by first identifying transients coincident with confirmed or candidate lens systems. We construct a catalogue of lens systems based on a compilation of literature sources covering a variety of surveys and identification methods \citep{bolton--08, more--12, sonnenfeld--13, gavazzi--14, cao--15b, more--16, diehl--17, shu--17, sonnenfeld--18,  wong--18, jacobs--19a, jacobs--19b, petrillo--19, canameras--20, cao--20, chan--20, huang--20, jaelani--20, li--20, sonnenfeld--20, wong--20, huang--21, li--21, rojas--21, savary--21, shu--22}. We also include all objects for which the convolutional neural networks (CNNs) presented by \cite{rojas--21} and \cite{savary--21} have assigned a $\geq0.5$ probability of being a lens ($\sim$133\,000 and $\sim$9\,200, respectively). In total, after removing duplicates and those outside the ZTF footprint, our compilation includes $\sim$60\,000 systems. We use a 5\farcs{} radius (which is typically the maximum image separation predicted from simulations of glSNe; \citealt{goldstein--19, wojtak--19}) for each lens to crossmatch against our transient sample. In total we find three transients coincident with a candidate lens: ZTF18aawhhey, ZTF21aacklas, and ZTF21aclrkdk.

\par

ZTF18aawhhey shows a relatively flat light curve just above the ZTF detection threshold and is offset by 1.9\farcs{} from a candidate lens presented by \cite{petrillo--19}. The candidate lens was ranked by \cite{petrillo--19} as having a high probability of being a lens, based on visual inspection (score = 20) and a CNN (score = 0.998). Using the photometric redshift code presented by \cite{tarrio--20}, we estimate the redshift for the candidate lens as $z = 0.26\pm0.02$. Assuming a peak magnitude of $m_{{r}} \leq 20.24\pm0.17$ (and no host extinction), correcting for the distance and Milky Way extinction ($A_{{V}} = 0.05$), this would imply a peak mag of $M_{{r}} \leq -20.41\pm0.26$. The poorly constrained light curve makes a definitive classification difficult, but we find it can also be fit by SNe~Ia templates at $z \sim 0.2$. Therefore we see no evidence in favour of being a glSN. ZTF21aacklas was offset by 0.24\farcs{} from a candidate lens identified by \cite{savary--21}, but shows stochastic or likely AGN variability. Our final candidate, ZTF21aclrkdk, is offset by 2.9\farcs{} from a candidate lens identified by \cite{rojas--21} at $z = 0.33\pm0.05$. A closer galaxy however at 1.5\farcs{} and $z = 0.15\pm0.10$ is likely the correct host. We find that ZTF21aclrkdk is consistent with a SN~Ia light curve at $z = 0.15\pm0.10$, although again the light curve is poorly constrained.

\par

As surveys push to deeper limits and catalogues of candidate lens systems become more complete, the probability for chance alignment increases. Assuming a radius of 5\farcs{} around each of our candidate lenses, this results in an effective sky area of $\sim$0.38~deg$^2$. The ZTF public survey has a total footprint of $\sim$23\,675~deg$^2$ \citep{bellm--19}, therefore our lens catalogue covers 0.002\% of the survey. Assuming each of the lenses in our catalogue and each of the 15\,215 transients in our sample are randomly distributed, this would indicate that we should find $\sim$0.2 crossmatched transients. Given that lenses and transients are typically not randomly distributed, but concentrated around galaxies, we consider this to be a lower limit and consistent with our three coincident sources.

\par

We also note that another transient, ZTF22abdyjqu, was discovered coincident with a candidate lens after the cut-off date for our sample. ZTF22abdyjqu was separated by $\sim$1.6\farcs{} from DESI-311.4249-10.6762, a B-grade candidate lens with a spectroscopic redshift of $z = 0.6334$ presented by \cite{huang--21}. Follow up observations for classification were triggered, but not completed due to weather. A classification spectrum was obtained by ePESSTO+ \citep{pessto} that confirmed ZTF22abdyjqu as an unrelated foreground SN~Ia at $z = 0.108$ \citep{ZTF22abdyjqu--class}.

%

\begin{figure*}
\centering
\includegraphics[width=\textwidth]{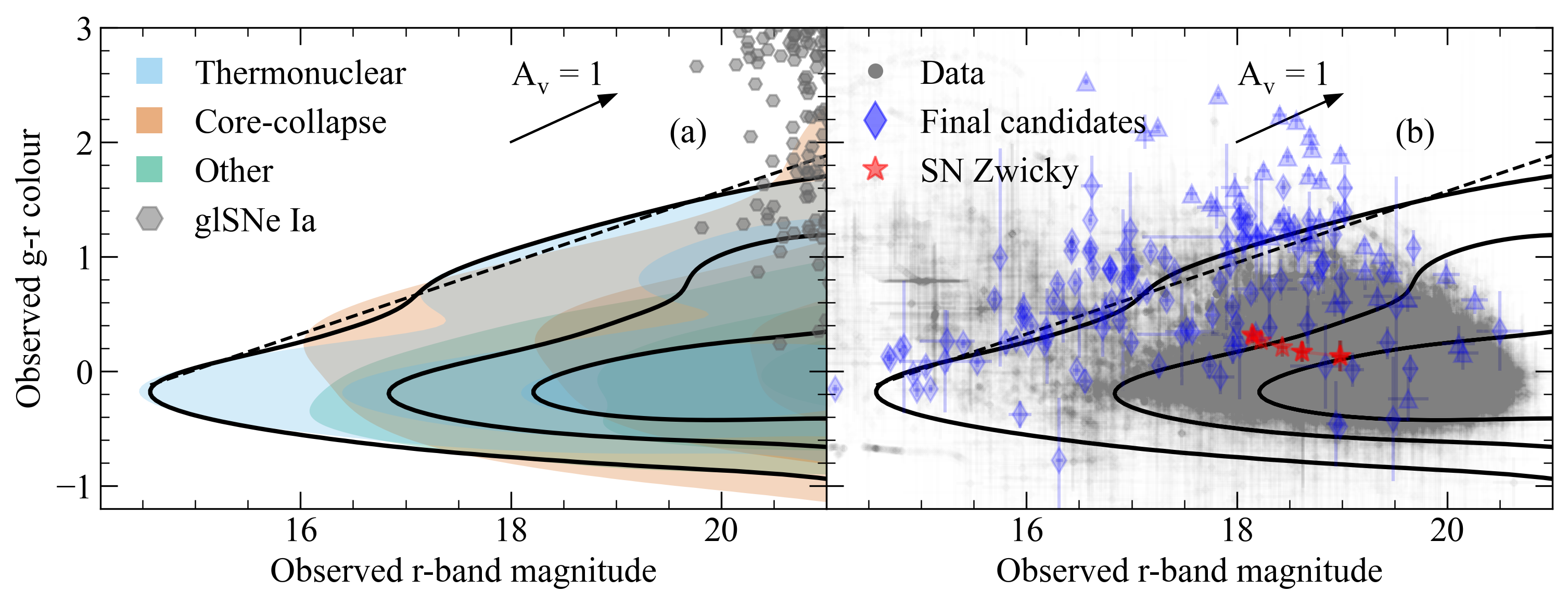}
\caption{\textit{Panel a: }Colour magnitude diagram for simulated non-lensed transients in the ZTF $g$- and $r$-bands. Transient light curves are generated from the beginning of each template up to maximum light with a 1~day cadence to provide a well-sampled rise. Shaded regions show the 1, 2, and 3$\sigma$ regions of the colour-magnitude space for each transient type. The thick black lines show the 1, 2, and 3$\sigma$ ranges across all types of transients. A linear fit to this limit is shown as a dashed line. The arrow shows the change in colour and magnitude expected for an extinction of A$_{V} = 1$. Simulated glSNe~Ia are shown as grey hexagons and discussed in Sect.~\ref{sect:colour_method_limits}. \textit{Panel b: }Distribution of colours and magnitudes from our observed sample of public ZTF transients. Observations of our final candidate sample are shown as purple diamonds. Lower limits on the transient colour, based on non-detections in the $g$-band, are shown as triangles. We note that only the first and last observation during the rise are shown here. SN~Zwicky is shown as a red star.
}
\label{fig:colour_method}
\centering
\end{figure*}

\section{Very red transients}
\label{sect:colour_method}

Our first selection method is based on the transient colour and follows from \cite{quimby--14}, motivated by the discovery of PS1-10afx. PS1-10afx was initially classified as a superluminous SN (SLSN), due to its high luminosity ($M_{u} \sim -22.3$) and spectral similarities to other SLSNe and SNe~Ic \citep{chornock--13}. In contrast, \cite{quimby--13} instead argue PS1-10afx is consistent with being an extremely high redshift ($z = 1.388$) SN~Ia that was magnified by a factor of $\sim$30. Subsequent observations of the host galaxy after the SN had faded revealed the presence of the lens galaxy -- indicating that PS1-10afx was indeed likely a gravitationally lensed SN~Ia \citep{quimby--14}.

\par

Although the initial classification of PS1-10afx was unclear, it immediately stood out from other SNe due to its extremely red colours, as the high redshift had caused most of the rest-frame optical flux to shift out of the observer-frame optical bands. Around maximum light, PS1-10afx showed an $r-i$ colour of $\sim$1.7~mag, compared to $r-i \textless 0.5$ for non-lensed SNe \citep{quimby--14}. Based on this observation, \cite{quimby--14} suggest a colour-based selection method for identifying glSNe. Using Monte-Carlo simulations, they estimate the expected colours for lensed and non-lensed SNe in PS1. As a function of apparent magnitude, they calculate the upper limit on the $r-i$ colour for non-lensed SNe as they rise towards maximum light and suggest selecting candidate glSNe if they are redder than this limit (see their fig.~4). Following \cite{quimby--14}, we apply this method to our search for glSNe in the ZTF filters.

\par

\subsection{Selection method}
To calculate the distribution of colours for non-lensed transients in the ZTF filters, we used the PLAsTiCC templates \citep{kessler--19} and include SNe~Ia, 91bg-likes, SNe~Iax, SNe~II, SNe~Ibc, SLSNe, Ca-rich, and tidal disruption events (TDEs). Following the methods outlined in \cite{kessler--19}, we generate random distributions of each type of transient based on the provided templates, up to a redshift of $z = 0.15$, and calculate light curves in the ZTF $g$- and $r$-bands during the rising phase up to $r$-band maximum. Rates for each type of transient were also taken from \cite{kessler--19} and increased by a factor of four. This ensures the relative numbers of transients are consistent with observations, while also better sampling the underlying distributions. Milky Way extinction is not included in our simulations as we assume this can easily be corrected for, while host galaxy extinction is also not included to limit the number of potential glSNe that are discarded. For our selected candidates, we investigate whether host galaxy extinction could have produced the observed red colours (Sect.~\ref{sect:colour_method_candidates}).

\par

Figure~\ref{fig:colour_method}(a) shows the distribution of colours and magnitudes for our simulated non-lensed SNe during the rising phase. As expected, different transient types show different colour ranges. The thermonuclear group (SNe~Ia, 91bg-likes, and SNe~Iax) shows a somewhat bimodal distribution, with SNe~Ia and SNe~Iax typically being relatively blue during their rise ($g-r \lesssim 0.2$). Conversely 91bg-like SNe are generally fainter and redder, with $g-r \gtrsim 0.5$ during the rising phase. Core-collapse SNe (SNe~II and SNe~Ibc) show a broad range of colours from $-0.5 \lesssim g-r \lesssim 1.0$. Finally, for other types of transients (SLSNe, TDEs, and Ca-rich), we also find a broad range of colours. This is mostly due to the Ca-rich transients as SLSNe and TDEs are typically blue during the rise. From these simulations, we use the Kernel Density Estimate from \textsc{seaborn} \citep{seaborn} to find the 1, 2, and 3$\sigma$ ranges of the colour-magnitude space across all simulated light curves and transient classes. These are shown in Fig.~\ref{fig:colour_method} as thick black lines. An approximate linear fit to the upper limit of the 3$\sigma$ distribution is shown by a dashed black line and given by the following functional form:
\begin{equation}
    g-r = 0.3116 \times m_{r} - 4.663.
\end{equation}

\par

For each of the observed 15\,215 transients in our sample, we fit their light curves using Gaussian Processes (GP) to estimate the colour evolution during the rising phase. GPs are commonly used to interpolate transient light curves (e.g. \citealt{inserra--13a,dhawan--18,narayan--18}). Here we use a combination of three squared-exponential kernels to account for variations of different timescales across transient populations. As we do not place a constraint on the number of $g$-band detections, in cases where there are no suitable $g$-band detections from which to measure a colour, we use the non-detections from the same night to set a lower limit on the colour. We note however that these non-detections do not necessarily imply extremely red colours and should be treated with caution. Non-detections are reported based on an average limiting magnitude across the whole image and not forced photometry at the site of the transient \citep{masci--19}. Therefore, while some non-detections in the $g$-band could be related to red colours, others may result from issues with the survey or data processing (for example, the transient falls on a chip gap or a bad galaxy subtraction). Colours for the transients in our sample are shown in Fig.~\ref{fig:colour_method}(b) as grey points. 

\par

We find 445 transients with $g-r$ colours and $r$-band magnitudes outside the 3$\sigma$ region predicted by our simulations, after correcting the colours and apparent magnitudes for Milky Way extinction. Visually inspecting all of the colour outlier candidates we find a large number were flagged as outliers based on showing rising features in the light cure, despite being clearly post-maximum. This includes SNe~Ia in which the rise around the secondary maximum was identified, SNe with minor fluctuations in the light curve causing multiple portions of the light curve to appear to rise, and some transients clearly showing repeating and/or stochastic variability similar to variable stars or AGN. Finally, we also find a number of cases of deep non-detection $g$-band limits during the transient evolution, implying very red colours, even when the $g$-band detections on earlier or later nights indicate the transient was relatively bright. Again, this could arise from issues with the survey data. Removing all of these cases and selecting only those transients that were flagged during an unambiguous rising phase leaves 84 in our sample. Applying even stricter cuts and selecting only those objects with colours more $\textgreater$3$\sigma$ outside our range would reduce the sample even further to 48 transients. To limit the number of potential glSNe that are discarded, in the following section we include the full sample of 84.

\par

\subsection{Candidates}
\label{sect:colour_method_candidates}
We identify 84 candidates for glSNe within our sample of 15\,215 transients, based on their red colours before $r$-band maximum. For each of these candidates, we query NED to identify potential host galaxies nearby with redshift estimates. For those transients without nearby galaxies catalogued on NED, we again use the PS1 photometric redshift code presented by \cite{tarrio--20} to estimate the redshifts of nearby galaxies. Using these known or estimated redshifts, we fit the light curves with various SN templates. This allows us to estimate whether the red colours could plausibly be explained by unlensed transients that are intrinsically red or experience varying degrees of extinction. The majority of these candidates (67) have been spectroscopically classified, therefore we also investigate whether these spectroscopic classifications are consistent with the observed light curves. 

\par

We find that candidates identified by our selection criteria generally fall into one of the following categories:

\begin{enumerate}
    \item Becoming brighter than the 3$\sigma$ range predicted by our simulations
    \item Red colours due to mis-estimated or poorly-sampled GP fits
    \item Red colours due to declining flux in the $g$-band
    \item Red colours due solely to non-detection limits
    \item Intrinsically red and/or heavily extincted light curves
\end{enumerate}

\begin{figure}
\centering
\includegraphics[width=\columnwidth]{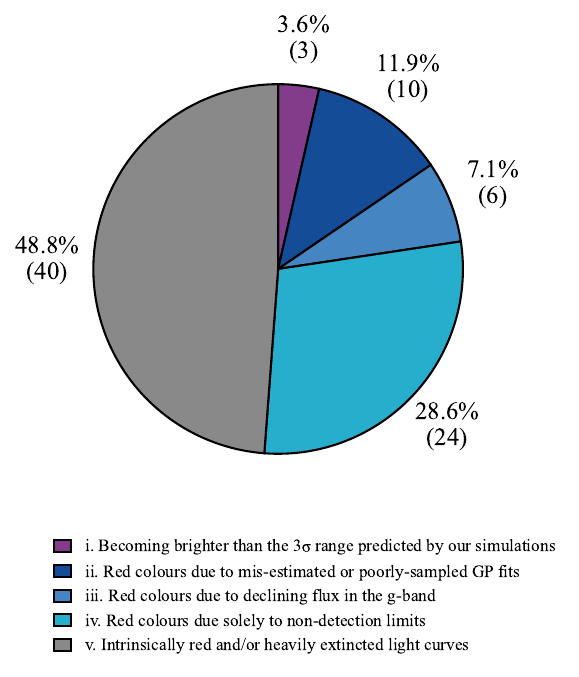}
\caption{Pie chart showing the number of transients falling into each outlier category for candidates identified via the very red transients method. }
\label{fig:colour_method_pie}
\centering
\end{figure}

We note that including only those objects that are 3$\sigma$ outliers would remove all transients in category ii. For some candidates selected solely from red colours estimated with non-detection limits, we cannot conclusively determine their classification or whether their red colours are intrinsic. Nevertheless, in all cases we find that the available $r$-band observations are consistent with existing SN templates at the redshift of the nearby galaxy. Figure~\ref{fig:colour_method_pie} shows a pie chart detailing the number of candidates flagged for each category. More detailed descriptions regarding each of the various sources of contamination are presented in the appendix in Sect.~\ref{sect:appdx:colour_outliers} along with a full list of our identified candidates in Table~~\ref{tab:colour_outliers}. Overall, we find all of our candidates are consistent with non-lensed transients. In Sect.~\ref{sect:limits} we discuss limitations of this method.

%

\section{Light curve inconsistent with SN~Ia at photometric redshift of nearby elliptical galaxy}
\label{sect:outlier_method}

\cite{goldstein--18--lens} suggest a method of identifying candidates based on model fits to their light curves and selecting those that are incompatible with being SNe~Ia at the apparent host galaxy redshift. They argue that most strong gravitational lenses in the Universe are massive elliptical galaxies and that most SNe that occur in elliptical galaxies are SNe~Ia. Given that most glSNe will be discovered in systems for which there are no lens arcs easily observed, the glSN will therefore appear to be hosted by the lens galaxy. Hence if a SN appears to be hosted by an elliptical galaxy, but is not a SN~Ia at that redshift, it could be a glSN.

\par

Using a series of SN templates, \cite{goldstein--19} generate ZTF light curves for glSNe. For each glSN they fit the light curve using the standard SALT2 model \citep{guy--07} at the redshift of the apparent host galaxy, in other words the strong gravitational lens in their simulations. They show that this results in significant outliers relative to the SALT2 model, due to the inconsistent redshift, and therefore they can reliably select candidate glSNe. We apply this same selection method to our sample of 15\,215 ZTF transients.

\subsection{Selection method}
\label{sect:outlier_method:select}
The method used by \cite{goldstein--19} requires selecting SNe that appear to be hosted by elliptical galaxies. \cite{goldstein--19} assume a complete catalogue of elliptical galaxies and hence all glSNe in their simulations pass this initial cut. In the present work, we do not have complete catalogue and therefore require some additional selection cuts. To identify elliptical galaxies, we use the selection criteria outlined by \cite{irani--22}, which were used to select core-collapse SNe observed by ZTF that occurred in elliptical galaxies. This method is based on photometry from the GALEX Data Release 8/9 (\citealt{martin--05}; FUV and NUV bands) and the ALLWISE catalogue (\citealt{wright--10}; W2 and W3 bands), in addition to the PS1 photometric catalogue DR2 ($r$-band). Following from \cite{irani--22}, if a galaxy satisfies either of the criteria given below, it is selected as likely being elliptical.
\begin{itemize}
    \item If photometry is available in all of the $W2$-, $W3$-, $NUV$-, and $r$-bands: $W2 - W3 \leq 0.5$ and $NUV -$ r $\geq$ 3.
    \item If photometry is available in only the $W2$ and $W3$ bands: $W2 - W3 \leq 0.3$.
\end{itemize}
For the objects in our sample, we query the required catalogues and find 1\,241 are within 5\farcs{} of an elliptical galaxy. To reduce the number of contaminants from non-ellipticals, in cases where only upper limits on the colour are available we include only those galaxies for which the upper limit is within the criteria outlined above. For 99\% of the objects selected here, the elliptical galaxy is the closest galaxy within 5\farcs{}. In addition, we find that 0.5\% of the objects in our sample are within 5\farcs{} of two elliptical galaxies, while the rest are coincident with one.

\par

Having identified the objects in our sample which could be hosted by elliptical galaxies, we now look to establish whether they are consistent with a SN~Ia at the required redshift. During their analysis, \cite{goldstein--19} also assume that each catalogued elliptical galaxy has a secure photometric redshift. For this work, we again use the photometric redshift code presented by \cite{tarrio--20} and photometry from the PS1 DR2 to estimate redshifts for each of the elliptical galaxies in our sample. Due to non-detections in multiple PS1 bands, we were unable to estimate redshifts for 38 potential hosts in our sample.

\par

For the remaining 1\,203 objects in our sample, we follow \cite{goldstein--18--lens} and fit the light curves of each object with the SALT2 template and the redshift of the potential elliptical host galaxy (or galaxies in the case of objects with more than one nearby elliptical galaxy). Unlike \cite{goldstein--18--lens}, who assumed a precise redshift was known, here we allow the redshift to vary across the 3$\sigma$ range predicted by the \cite{tarrio--20} photometric redshift code. As in \cite{goldstein--18--lens}, we also require $|x_1| \leq 1$, $|c| \leq 0.2$, and a peak magnitude within the range expected for normal SNe~Ia ($ -18.5 \leq M_g \leq -20$). Using this method we find 310 objects in our sample have at least one 5$\sigma$ outlier from the best-fit SALT2 model across the 3$\sigma$ redshift range.

\par 

Following visual inspection of our model outlier candidates, we again find multiple objects with long, repeating, and/or stochastic variability consistent with stellar variables or AGN. Removing each of these objects leaves 133 in our sample. As in Sect.~\ref{sect:colour_method}, a number of candidates were flagged solely on the basis of deep, non-detection limits that were inconsistent with the rest of the light curve evolution. Removing these candidates leaves 73 in our sample. We also find a number of candidates were flagged as outliers due to likely spurious or bogus detections. We note that each detection from \alerce{} also includes a real/bogus score to identify spurious detections, however we found a number of cases where clearly bogus detections were flagged as real and vice-versa, and therefore we choose not to implement a cut based on this score. We rely on visual inspection to remove objects that were identified as candidates based on spurious outliers. This leaves 56 in our sample. For 22 candidates, we find their light curves are consistent with normal, but well-observed SNe~Ia. In these cases, the observations extend to phases beyond the temporal coverage of the SALT2 template (i.e. $\textgreater+50\,d$) and therefore were flagged as outliers. Removing these observations and fitting only the light curve within $-15$\,d $\leq t_0 \leq 50$\,d we find no outliers for any candidate and therefore remove them, leaving 34 in our sample. Finally, we visually inspect the host galaxies of each of the remaining candidates and remove any objects for which spiral arms are clearly identifiable in archival PS1 or SDSS imaging, leaving 29 candidates.

\subsection{Candidates}
We identify 29 candidates for glSNe within our sample of 15\,215 transients, based on having at least one 5$\sigma$ outlier relative to a SALT2 fit at the redshift of the nearby elliptical galaxy. We follow a similar method as in Sect.~\ref{sect:colour_method_candidates} and fit the light curve of each transient with various SN templates to determine whether they are consistent with being unlensed SNe. Again, we find that the majority (16) of our candidates have been spectroscopically classified (11 of which are SNe~Ia).

\par

As in Sect.~\ref{sect:colour_method_candidates}, we find that our selected candidates can be grouped into a handful of categories:

\begin{enumerate}
    \item Normal SNe~Ia
    \item Peculiar SNe~Ia
    \item Other transients
    \item Light curves with multiple peaks
    \item Incorrect host redshifts
\end{enumerate}

\begin{figure}
\centering
\includegraphics[width=\columnwidth]{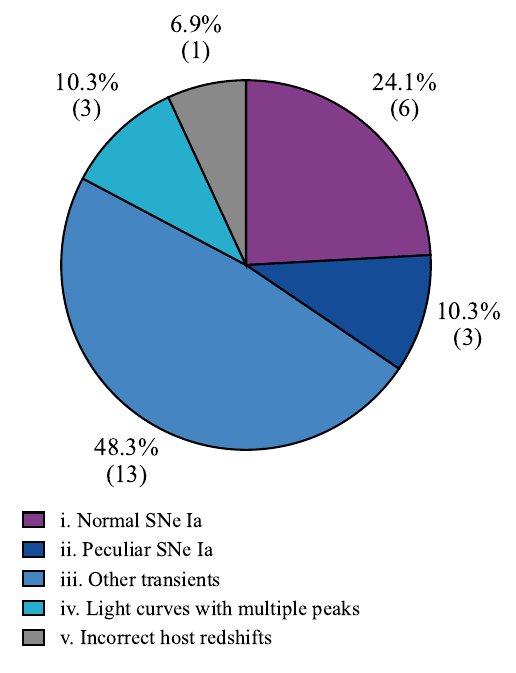}
\caption{Pie chart showing the number of transients falling into each outlier category for candidates identified via a light curve that is inconsistent with being a SN~Ia at the photometric redshift of the nearby elliptical galaxy. }
\label{fig:model_outlier_method_pie}
\centering
\end{figure}

Figure~\ref{fig:model_outlier_method_pie} shows a pie chat with a breakdown of the number of candidates falling into each outlier category. Detailed descriptions of each of these contamination sources are given in the appendix in Sect.~\ref{sect:appdx:model_outliers} along with a complete list of the selected candidates in Table~\ref{tab:model_outliers}. For some candidates we are unable to provide a definitive classification, nevertheless all light curves are consistent with low-redshift transients and we find no evidence in support of glSNe. Limitations of this method are discussed in Sect.~\ref{sect:outlier_method_limits}.

%

\section{Intrinsically luminous assuming host redshift}
\label{sect:luminosity_method}
Our final method of searching for glSNe is based on the discovery of iPTF16geu \citep{goobar--17} and SN~Zwicky \citep{goobar--22} from unresolved, lensed images. Both SNe became sufficiently bright to trigger spectroscopic classification, which showed them to be high redshift SNe that had been highly magnified. Spectra of iPTF16geu showed it to be consistent with a SN~Ia at $z = 0.41$, however the peak apparent magnitude of $m_B = 19.12$ implied an absolute magnitude of $M_B \lesssim -22.48$. For SN~Zwicky, spectra showed a SN~Ia at $z = 0.35$ and peak apparent magnitude of $m_B = 18.49$, implying an absolute magnitude of $M_B \lesssim -22.78$. Such high peak luminosities are not expected for SNe~Ia and therefore indicated that both objects had been significantly magnified. Here, we look for similar objects with inferred bright absolute magnitudes among our sample of 15\,215 transients.

\par

\subsection{Selection method}
To identify transients that show evidence of bright absolute magnitudes we first apply the directional light radius method \citep{sullivan--06, gupta--16} to select the most likely host galaxy within 5\farcs{} from PS1 for each object in our sample. We find 11\,056 transients with a PS1 host galaxy within 5\farcs{}. We again use the photometric redshift code from \cite{tarrio--20} and photometry from the PS1 DR2 to estimate redshifts. To reduce contamination, we require a well-constrained photometric redshift whereby $\frac{\sigma_z}{1+z} \leq 0.05$ \citep{goldstein--17}, giving 7\,586 in our sample. Selecting the brightest detection in the light curve, we assume the 3$\sigma$ lower bound on the photometric redshift of the host galaxy and observed magnitude to identify any transient showing evidence of an absolute magnitude $M_r \leq -21$ (typically used as a cut-off to identify SLSNe; \citealt{gal-yam--12}). Applying this method, we find 28 transients with inferred absolute magnitudes $M_r \leq -21$. Following visual inspection, we remove any transients with long, repeating and/or stochastic variability, leaving 19 candidates. Separately, we also follow a similar method using SDSS spectroscopic host redshifts. In this case we find one transient, ZTF20aaaweke, with $M_r \leq -21$, which was spectroscopically classified as a SN~IIn at the redshift of the host galaxy \citep{ZTF20aaaweke--class}.

\par

\subsection{Candidates}
Based on the PS1 photometric redshift of the likely host galaxy, we identify 19 candidate glSNe with inferred bright absolute magnitudes. These selected candidates are given in Table~\ref{tab:bright_outliers}. Four of these candidates were spectroscopically classified as low-redshift SNe~Ia and selected due to incorrect host association. In all cases, we find galaxies with photometric redshifts consistent with the spectroscopic observations at larger separations. One candidate was spectroscopically classified as a high-redshift TDE.

\par

For the remaining 14 candidates, the limited data and data quality mean we are unable to provide a definitive classification however we note that most are associated with faint galaxies that show no evidence of lensing features. In all cases, we find the transient light curves are consistent with SLSN templates \citep{kessler--19}. We also attempt to fit their light curves using the SLSN module in \textsc{mosfit} \citep{mosfit}, however these fits are unable to converge for most candidates. Assuming the bright absolute magnitudes are due to incorrect photometric redshift estimation, we fit the light curves using the SALT2 model assuming $x_1 = c = 0$ (to reduce degeneracy) and allowing the redshift to vary freely. In most cases we find these SALT2 models are able to reproduce the observations and the resulting redshifts are also given in Table~\ref{tab:bright_outliers}. Based on the data available, we do not find significant evidence in favour of glSNe among this sample, however we again note definitive classifications are not possible.

%

%

\section{Discussion}
\label{sect:discuss}

\subsection{Combined methods}
\label{sect:combined}
Each of the search methods outlined in previous sections were applied to our sample independently. Combining them however may be able to identify the most promising candidates for glSNe -- those that are close to elliptical galaxies and show red colours with high inferred peak luminosities. Assuming either SDSS spectroscopic or PS1 photometric redshifts, no candidate was selected via all three methods.

\par

Combining the 445 candidates selected based on red colours, the 310 selected based on outliers relative to a SALT2 model, and the 28 selected based on bright absolute magnitudes, we find 104 candidates were selected by two of these methods using the initial set of selection criteria. In total, we find 132 unique transients that were identified by the final set of selection criteria in any of our methods, with none being ultimately selected by multiple methods.

\subsection{Limitations}
\label{sect:limits}
Here we discuss the limitations of the methods used during this work to identify candidate glSNe. Although discovered after the cut-off date for our initial sample, SN~Zwicky provides the perfect opportunity of testing the different identification methods used here and whether they would have been able to successfully identify it.

\subsubsection{Very red transients}
\label{sect:colour_method_limits}
With a relatively low redshift ($z = 0.35$), SN~Zwicky did not show particularly red colours during the rising phase. Using the publicly available alert photometry from \alerce{}, and correcting for Milky Way extinction, SN~Zwicky increased from $m_r \sim$19.0 -- 18.2 following discovery. Throughout this period, the $g-r$ colour became redder, from $\sim$0.1 -- 0.3. As shown by Fig.~\ref{fig:colour_method}, SN~Zwicky would therefore not have passed our colour cuts and would be excluded. Including only SNe~Ia templates in our colour calculation however, SN~Zwicky would have been selected around maximum, but this would also lead to significantly increased contamination from other SN types.

\par

The primary limitation of the colour-based method applied here is that, for relatively shallow surveys such as ZTF, extreme magnifications are required. Therefore there may be many glSNe that simply do not reach the required apparent magnitude to be recognised as outliers. To estimate the fraction of glSNe that would be sufficiently magnified to pass our colour-based selection criteria, we simulate a sample of lensed and unlensed SNe~Ia. 

\par

The rate of glSNe that will be detectable by ZTF is given by
\begin{equation}
    \frac{dN_{SL}}{d{z_s}}(z_s) = \frac{dN_s(z_s)}{dz_s}\tau(z_s)B(z_s).
\end{equation}
Here, $\frac{dN_s}{dz_s}$ corresponds to the SN~Ia rate as a function of redshift. As in Sect.~\ref{sect:colour_method}, we use the rate given by \cite{kessler--19}. The factor $\tau(z_s)$ is the lensing optical depth and represents the probability of a source at redshift $z_s$ being strongly lensed. Finally, $B(z_s)$ accounts for the magnification bias, which is the fact that glSNe will be drawn from a fainter source population than unlensed SNe due to their higher redshifts.

\par

 Assuming the mass profile of lens galaxies is given by a singular isothermal sphere (SIS; \citealt{schneider-book}), the lensing optical depth is defined as
\begin{equation}
\tau (z_s) = \int_0^{z_s} dz_l \frac{d^2 V}{dz_l d\Omega } \int_{\sigma_{min}}^{\sigma_{max}} d\sigma \frac{dn}{ d\sigma } \pi [\theta_{\rm{Ein}} ( z_s,z_l,\sigma )]^2,
\end{equation}
where $V$ is the comoving volume element, $\frac{dn}{ d\sigma }$ is the number density of lenses, and $\pi$ [\rein $(z_s,z_l,\sigma )$]$^2$ is the lensing cross-section. We take the velocity dispersion function derived by \cite{bernardi--10} from SDSS DR6 for all galaxy types,
\begin{equation}
dn=\phi_{*} \left(\frac{\sigma}{\sigma_*}\right)^{\alpha}exp\left[ -\frac{\sigma}{\sigma_*}\right]^{\beta} \frac{\beta}{\Gamma(\alpha/\beta})\frac{d\sigma}{\sigma},
\end{equation}
where $\phi_{*}=2.099\times10^{-2}$~$(h/0.7)^3$~Mpc$^{-3}$, $\sigma_*=113.78$~km~s$^{-1}$, $\alpha=0.94$, $\beta=1.85$. Following from \cite{Collett_2015}, we assume a population of lenses with velocity dispersions $\sigma$ \textgreater 100~km~s$^{-1}$.

\par

The magnification bias, $B(z_s)$, is defined as
\begin{equation}
B(z_s)=\int_{\mu_{\text{Q}}(z_s)}^{\infty} d\mu P(\mu)W(f(\mu),f_{\text{lim}}).
\end{equation}
The minimum magnification required for a glSN to be detected by our colour-based method (i.e. to equal the black line in Fig.~\ref{fig:colour_method}) is given by $\mu_{\rm{Q}}(z_s)$. Following from \cite{schneider-book}, for the SIS a SN located at a distance $r$ from the centre of the source plane will have a magnification of the brightest lensed image given by
\begin{equation}
    \mu=\frac{r+1}{r}.
    \label{eqn:muSIS}
\end{equation} 
The probability of obtaining a given magnification is therefore related to the probability of a given position on the source plane. Here, $r$ is given in units of the Einstein radius and, for the case of strong lensing, multiple images are formed only when $r \leq 1$. Following from Eqn.~\ref{eqn:muSIS},\, $r = \frac{1}{\mu-1}$. Assuming a uniform distribution of sources across the source plane, such that $P(r) = 2r$ and $P(\phi) = \frac{1}{2\pi}$, the likely magnification distribution is given by 
\begin{equation}
    \int_0^R P(r) dr = \int^{\infty}_2 \frac{2}{(\mu - 1)^3} d\mu = \int^{\infty}_2 P(\mu) d\mu.
\end{equation}
$W(f(\mu),f_{\rm{lim}})$ gives the probability of a background source at $z_s$ being sufficiently magnified such that it would be detectable above the limiting flux of ZTF, $f_{\rm{lim}}$. Following \cite{collett--12}, $W$ is given by
\begin{equation}
W(f(\mu),f_{\rm{lim}})~=~ 
  \begin{cases}
   1  & \text{if } f(\mu) \geq f_{\text{lim}}/2 \\
   \left({2f(\mu)\over f_{\text{lim}}}\right)^2      & \text{if } f(\mu) < f_{\text{lim}}/2,
  \end{cases}
\label{eqn:w}
\end{equation}
where $f(\mu)$ is the magnified flux of the source. We assume a characteristic detection depth of $m_r = 20.5$ for ZTF \citep{bellm--19}. 

\par

To compute $B(z_s)$ we use the \cite{hsiao--07} spectroscopic template to simulate $10^4$ SNe~Ia light curves between $-14$ -- 0\,d relative to $B$-band maximum in a series of redshift bins up to $z = 2.5$. We chose the \cite{hsiao--07} spectroscopic template for this purpose as it extends to shorter wavelengths than other templates, which is necessary for simulating high redshift SNe. The peak $B$-band absolute magnitude of each simulated SN~Ia is drawn from a Gaussian distribution, $\mathcal{N}(-19.3,\,0.5)$. As in \cite{feindt--19}, host extinction is drawn from an exponential distribution with a rate of $\lambda = 0.11$. Within each redshift bin, we take the average of the $g-r$ colours and $r$-band magnitudes across all simulated SNe~Ia to compute the minimum magnification, $\mu_{\rm{Q}}(z_s)$.

\par

In Fig.~\ref{fig:colour_method} we show a sample of simulated glSNe (from a total of 50\,000) calculated using the above method. The vast majority of these glSNe will not be sufficiently magnified to be detected via this method and many fall below our colour upper limit. This was also observed by \cite{quimby--14} in their simulations (see their figure 4).

\par

\begin{figure}
\centering
\includegraphics[width=\columnwidth]{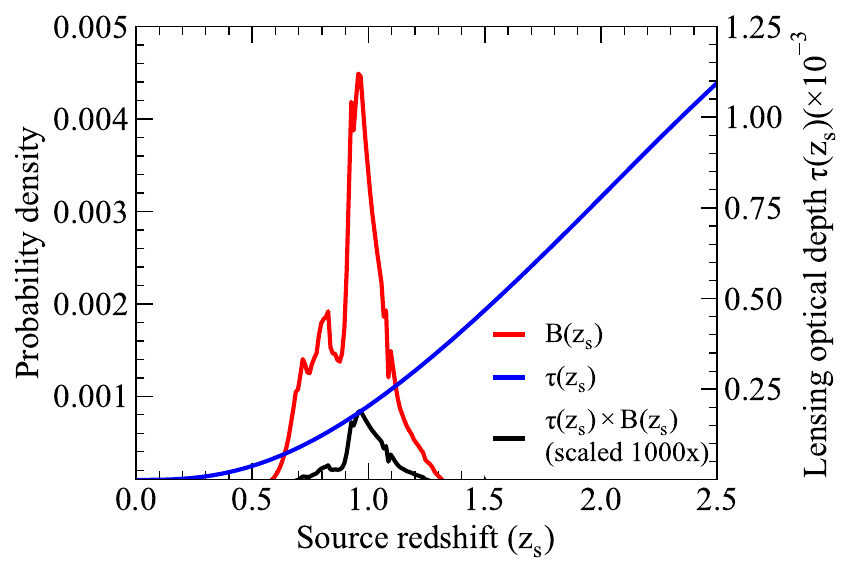}
\caption{Magnification bias, $B(z_s)$ (left axis), and lensing optical depth, $\tau(z_s)$ (right axis), as a function source redshift, $z_s$. The product $\tau(z_s) \times B(z_s)$ gives the total probability that any individual SN is strongly lensed and magnified such that it would be detectable by our colour-based method and above the flux limit of ZTF. We note that $\tau(z_s) \times B(z_s)$ is shown scaled up by a factor of $1\,000\times$.  }
\label{fig:lensedprobability}
\centering
\end{figure}

 The total probability that a glSN could be detected with this method, as a function of redshift, is given by the product of the magnification bias and lensing optical depth for each redshift bin and is shown by Fig.~\ref{fig:lensedprobability}. Figure~\ref{fig:lensedprobability} highlights that this selection method is only sensitive to glSNe within the redshift range $0.6 \lesssim z_s \lesssim 1.2$ and with an overall low probability of $\sim$$10^{-7}$ -- $10^{-6}$. This redshift range is slightly higher than that found by \cite{goldstein--19} using their method (Sect.~\ref{sect:outlier_method}). This likely results from the fact that only very high redshift SNe will be sufficiently red such that they lie outside the 3$\sigma$ range of our simulations when highly magnified.

\par

While the overall probability of detecting glSNe is low, to accurately evaluate the efficiency of this method we must compare to the number of non-lensed SNe~Ia that are also detectable. In other words, what is the rate of glSNe~Ia identified relative to non-lensed SNe~Ia that ZTF could actually discover? The flux limit of ZTF will only be sensitive to SNe~Ia with $z_s \lesssim 0.15$. Based on our simulations and the SN~Ia rate from \cite{kessler--19}, we find $\sim$0.06\% of SNe~Ia up to $z_s = 2.5$ would reach the flux limit of ZTF. Therefore, comparing the probability of being strongly lensed and passing our selection cuts ($\sim10^{-7}$ -- $10^{-6}$) to the probability of any given SN~Ia over the same redshift range meeting the flux limit of ZTF ($\sim$10$^{-3}$), we find an overall expected rate for detectable glSNe~Ia of $\sim$10$^{-4}$ per detectable unlensed SN~Ia.

\par

Based on our expected rate of glSNe~Ia ($\sim$10$^{-4}$), we can estimate the absolute number that should be detectable by ZTF. Among our sample of 15\,215 transients, using the \alerce{} light curve classifier \citep{sanchez-saez--21} we find that $\sim$10\% of objects are classified as not having `transient' light curves and therefore could represent contamination due to AGN or stochastic variability. We note however that in Sect.~\ref{sect:colour_method} \& \ref{sect:outlier_method}, the contamination rate was significantly higher for our identified candidates. Assuming $\sim$10 -- 50\% contamination from non-SNe results in a sample of 7\,608 -- 13\,694 potential SNe. In addition, from our sample of 15\,215 transients, 4\,157 were spectroscopically classified with 2\,877 being SNe~Ia. This fraction of SNe~Ia relative to all SNe ($\sim70\%$) is typical for a magnitude-limited sample \citep{li--2011}. Therefore assuming a similar fraction across our whole sample would result in $\sim$5\,326 -- 9\,586 potential SNe~Ia and imply we expect $\sim$0.5 -- 1.0 glSNe~Ia to have been detected by this method. We note that the simulations discussed here include only SNe~Ia. We expect that the fainter peak magnitudes of core-collapse SNe would likely result in an even lower probability of passing our selection method for this class, due to the larger magnifications required to reach the flux limit. Although the contamination rate is highly uncertain, our simulations show that we naively expect $\lesssim$1 glSN to be detected by this method. Therefore it is unsurprising that our search yielded no positive detections of glSNe, despite including four years of observations.

\par

\subsubsection{Light curve inconsistent with SN~Ia at photometric redshift of nearby elliptical galaxy}
\label{sect:outlier_method_limits}
SN~Zwicky is offset by 1.25\farcs{} from the host galaxy, which is present in the ALLWISE catalogue with a colour upper limit of $W2 - W3 \leq 1.89$. SN~Zwicky therefore would not have passed our selection criteria outlined in Sect.~\ref{sect:outlier_method:select}, however we note that we cannot conclusively rule out the colour as being consistent based on an upper limit. 

\par

Even allowing the galaxy to pass this initial selection cut, on the basis of the colour being an upper limit, SN~Zwicky would still not have been flagged as a candidate given that the host galaxy is visible and the lens galaxy is not. In other words, the nearby elliptical galaxy is the host, not the lens. Therefore fitting a SALT2 model to SN~Zwicky with the redshift of the elliptical would not have produced any significant outliers given that the redshift is correct. Indeed, we find a photometric redshift of the host of $z = 0.41\pm0.11$, which is consistent with the spectroscopic redshift of the SN, and no significant outliers when fitting the light curve with SALT2. Including this information on the host redshift and therefore luminosity, as in Sect.~\ref{sect:luminosity_method}, could have allowed SN~Zwicky to be recognised earlier, before the classification spectrum was observed. As discussed in Sect.~\ref{sect:luminosity_method} however, the use of photometric redshifts significantly increases the number of contaminants and the photometric redshift was not sufficiently well-constrained to pass our cuts. No prior spectroscopic redshift of the host was available.

\par

As discussed in Sect.~\ref{sect:outlier_method}, \cite{goldstein--19} assume a complete catalogue of elliptical galaxies. For this work, as we do not have a complete catalogue, we apply the selection cuts outlined in \cite{irani--22} to find elliptical galaxies coincident with transients in our sample. As discussed by \cite{irani--22}, these cuts are not 100\% complete, but are satisfied by 75\% of elliptical galaxies in Galaxy Zoo \citep{lintott--11}. Therefore, our selection of transients coincident with elliptical galaxies is likely at most 75\% complete and some glSNe could have been discarded on this basis.

\par

\cite{goldstein--19} also assume that the redshifts of elliptical galaxies in their catalogue are known exactly, which vastly reduces the degeneracy between SALT2 model parameters when fitting the observed light curves, particularly when fitting only two bands. For this work, we again do not have precise redshifts for each of the elliptical galaxies in our sample and therefore rely on photometric redshifts. In some cases, the uncertainty on the photometric redshift of the elliptical galaxy (i.e. the potential lens galaxy) may also be sufficiently large that the 3$\sigma$ range would also cover the redshift of any potential background glSN. Therefore in those cases no significant outliers would be flagged because the range of redshifts covered during the fit also includes the true value.

\par

Using this outlier-based detection method, \cite{goldstein--19} estimate that ZTF should discover 8.60 glSNe per year. From their simulations however, only $\sim$10 -- 16\% of these are detectable with data from the public survey alone. Therefore, across the four years of observations covered by our sample, we may expect to find 3.44 -- 5.50 glSNe. Assuming again that 25\% of our sample would not pass our elliptical selection criteria, this further reduces the number of glSNe to 2.58 -- 4.13. Given that we did not find any glSNe, this is within 1.6 -- 2.0$\sigma$ of the expected number from \cite{goldstein--19}. It is unclear however, how many glSNe would be discarded due to being fit with a redshift that does not produce any significant outliers, either due to degeneracies with other parameters (such as $x_1$ and $c$) or because the 3$\sigma$ range on the redshift of the lens galaxy would also include the redshift of the source. Therefore we consider these to be upper limits on the number of glSNe that should have been detected with this method and hence it is unsurprising that we were not able to confirm any positive detections.

\subsection{Estimating Einstein radii}

The methods for identifying glSNe used throughout this work have all resulted in significant numbers of contaminants, for which visually inspecting and performing classification is non-trivial. In the era of LSST, the number of SNe discovered will increase dramatically, making visual inspection of all candidates infeasible. Additional methods are therefore required to reduce the number of contaminants. One possibility is through estimating the Einstein radius of potential lens galaxies.

\par

The Einstein radius ($\theta_{\rm{Ein}}$) represents the separation between images of the lensed background source and the centre of the lens, and depends on the mass of the lens galaxy (or velocity dispersion) and distances between the observer, lens, and source. If a transient source is separated from nearby galaxies by significantly more than their Einstein radii, it is unlikely to have been strongly lensed and therefore may be rejected as a candidate glSN. For each of the 15\,215 transients in our sample, we estimate the Einstein radius of nearby galaxies and discard objects that are sufficiently separated, $\Delta \textgreater 3\theta_{\rm{Ein}}$.

\par

We estimate the most likely Einstein radius for all galaxies within 30\farcs{} of the transients in our sample. Again using the photometric redshift code presented by \cite{tarrio--20} and photometry from PS1 DR2, we calculate the absolute magnitude $M_{{r}}$ of each galaxy. Following from \cite{hyde--09}, we estimate the velocity dispersion, $\sigma$, as
\begin{equation}
    \log_{10} \sigma = \frac{-3 M_{{r}}^2 - 185M_{{r}} -1485}{500}.
    \label{eqn:rate}
\end{equation}
In addition to the velocity dispersion, the Einstein radius also depends on the distance to the source. We therefore assume any lensed background source could have a redshift up to $z = 2$ and calculate the angular diameter distances between the host and source, and observer and source in a series of redshift bins. The Einstein radius for each redshift bin $i$ is therefore given as
\begin{equation}
    \theta_{\rm{Ein},i} = 4 \pi \left(\frac{\sigma}{c}\right)^2 \frac{D_{\rm{ls},i}}{D_{\rm{s},i}},
    \label{eqn:rate}
\end{equation}
where $D_{\rm{s},i}$ is the angular diameter distance for a source in the current redshift bin and $D_{\rm{ls},i}$ is the angular diameter distance between the source and host galaxy. To calculate the most likely Einstein radius, we weight each one by the probability of strong lensing, $\theta_{\rm{Ein}}^2 W$. Here, $W$ is defined as in Eqn.~\ref{eqn:w} using the unmagnified peak magnitude of the background source (in this case $M_B = -19.3$ for an unlensed SN~Ia) and again the detection limit of ZTF.

\par

Calculating the most likely Einstein radii for galaxies coincident with each of our 15\,215 transients, we find 4\,241 are within 3\rein of any nearby galaxy. Applying this selection cut to our sample of 445 candidates identified based on red colours would have reduced this to 105 candidates. Taking only those that also passed our full set of cuts would have reduced the sample from 84 to 12. For those identified based on light curves inconsistent with being SNe~Ia at the photometric redshift of the nearby elliptical, our initial sample would have decreased from 310 to 209 while the final sample would decrease only slightly from 29 to 20. Finally, applying this selection cut to those transients selected based on bright absolute magnitudes would have removed all candidates. While ultimately all of our candidates were excluded from being glSNe, applying a cut based on the most likely Einstein radius could be an effective way of removing large numbers of contaminants. An additional cut on the minimum separation could also prove effective at removing AGN, which are the dominant source of contamination.

%

\section{Conclusions}
\label{sect:conclusions}
To date, only a handful of gravitationally lensed supernovae (glSNe) are known. We presented an extensive search for glSNe that may have previously been unclassified within four years of observations from the Zwicky Transient Facility (ZTF) public survey.

\par

We conducted an initial search by crossmatching our transient sample against a catalogue of $\sim$154\,000 known or candidate lens galaxies, compiled from various literature sources, and found three coincident sources within 5\farcs{}. All candidates however were likely due to chance alignment with a gravitational lens system.

\par

Using methods for finding glSNe that have been suggested in the literature (based on simulations), we also conducted a search for transients that were magnified by unknown lenses. Following the colour-based method outlined by \cite{quimby--14}, we performed simulations of unlensed SNe to estimate the range of colours produced during their rising phases.  Transients redder than this limit may be at high redshift and magnified to appear brighter, and are therefore candidate glSNe. Applying this method to our ZTF sample, we found 445 candidates. Visually inspecting each of these candidates, we rejected any that were clearly not pre-maximum (i.e. rising), leaving 84. Fitting each of these light curves, we found they were all consistent with existing templates for unlensed SNe. We therefore found no compelling evidence for candidate glSNe based on extremely red colours and bright magnitudes.

\par

The second detection method we implemented was outlined by \cite{goldstein--18--lens}. We identified all transients in our sample within 5\farcs{} of an elliptical galaxy and tested whether they were consistent with being SNe~Ia at the redshift of the elliptical. We found 310 candidates with at least one significant outlier relative to a SALT2 fit. Analysing each of these candidates, we found the main sources of contamination were due to bogus detections, well-observed SNe~Ia extending to phases beyond the SALT2 model, or AGN- or stellar-like variability. In addition, we also found two possible candidates for sibling SNe -- multiple SNe hosted by the same galaxy that are tagged as one object in survey data. Removing these contaminants, we again fit each light curve with existing SN templates and found no compelling evidence for glSNe. Finally, we also implemented a luminosity-based method to identify any transients with absolute magnitudes brighter than expected for normal SNe. Using photometric redshifts from PS1, we found 19 candidates, some of which were spectroscopically classified and selected due to incorrect host redshifts. For the remainder we are unable to present definitive classifications, but the observed light curves are generally consistent with being SLSNe or SNe~Ia at lower redshift than the PS1 photo-$z$. Using spectroscopic redshifts from SDSS, we identified one candidate, which was spectroscopically classified as a SLSN.

\par

To date, SN~Zwicky is the only confirmed glSN detected within ZTF. Although not included in our initial sample, we tested whether any of the methods used within this work would have successfully identified SN~Zwicky and found it would have been excluded in all cases. We also estimate the numbers of glSNe expected to be detected via these methods and find they are consistent with the null detections reported here. 

\par

In the coming years, LSST will discover thousands of SNe and should be more sensitive to glSNe given the increased observing depth and broader wavelength coverage \citep{goldstein--19}. The sheer volume of transients discovered by LSST mean that even with the relatively low efficiency rate of the methods used here, some glSNe should be discovered. Visual inspection or spectroscopic classification (as was the case for SN~Zwicky) for all candidate glSNe however will not be feasible. Therefore, given the false positive rate, the methods applied here to ZTF will not scale well to LSST.

\par

Alternatively, we suggest three complimentary approaches that would substantially improve the prospects of discovering lensed SNe in LSST. As most of the sky is not strongly lensed, an estimate of the likely Einstein radii for all elliptical galaxies would allow us to exclude most of the false positives. Watchlist functionality is already included in transient brokers and a simple watchlist of all elliptical galaxies, with specific association radii determined from the likely Einstein radii would significantly reduce contamination. One of the primary limitations of this work is the use of photometric redshifts which produce very uncertain Einstein radius estimates. A larger spectroscopic redshift catalogue compared to SDSS and a precise photometric redshift catalogue would allow us to more accurately find SNe that are much brighter than expected given the redshift of their host. Surveys and facilities such as DESI and 4MOST will provide spectroscopic redshifts for millions of galaxies and a more complete catalogue across a wider area. Finally, searches for and confirmation of galaxy-galaxy lenses should continue. A larger list of known lenses would likely provide the highest purity sample of candidates.

\par

This work has shown that pushing fainter than the very bright iPTF16geu and SN Zwicky-like lensed SNe is not trivial, even with full archival light curves. More work is needed to improve selection methods if real-time searches for cosmologically useful lensed SNe are to avoid being swamped by false positives.

\section*{Acknowledgements}

We thank R. Quimby for providing data on simulated glSNe. We thank the referee for their constructive comments.

This work has received funding from the European Research Council
(ERC) under the European Union’s Horizon 2020 research and in-
novation programme (LensEra: grant agreement No 945536). MRM acknowledges a Warwick Astrophysics prize post-doctoral fellowship made possible thanks to a generous philanthropic donation. TEC is supported by a Royal Society University Research Fellowship.

This work is licensed under a CC-BY license.

Based on observations obtained with the Samuel Oschin 48-inch Telescope at the Palomar Observatory as part of the Zwicky
Transient Facility project. ZTF is supported by the National Science Foundation under Grant No. AST-1440341 and a
collaboration including Caltech, IPAC, the Weizmann Institute for Science, the Oskar Klein Center at Stockholm University, the
University of Maryland, the University of Washington, Deutsches Elektronen-Synchrotron and Humboldt University, Los Alamos
National Laboratories, the TANGO Consortium of Taiwan, the University of Wisconsin at Milwaukee, and Lawrence Berkeley
National Laboratories. Operations are conducted by COO, IPAC, and UW. Based on observations obtained with the Samuel Oschin Telescope 48-inch and the 60-inch Telescope at the Palomar
Observatory as part of the Zwicky Transient Facility project. ZTF is supported by the National Science Foundation under Grants
No. AST-1440341 and AST-2034437 and a collaboration including current partners Caltech, IPAC, the Weizmann Institute for
Science, the Oskar Klein Center at Stockholm University, the University of Maryland, Deutsches Elektronen-Synchrotron and
Humboldt University, the TANGO Consortium of Taiwan, the University of Wisconsin at Milwaukee, Trinity College Dublin,
Lawrence Livermore National Laboratories, IN2P3, University of Warwick, Ruhr University Bochum, Northwestern University and
former partners the University of Washington, Los Alamos National Laboratories, and Lawrence Berkeley National Laboratories.
Operations are conducted by COO, IPAC, and UW.

This research has made use of the NASA/IPAC Extragalactic Database, which is funded by the National Aeronautics and Space Administration and operated by the California Institute of Technology.

\section*{Data Availability}

All ZTF data used in this work is publicly available and was accessed from the \alerce{} broker.



\bibliographystyle{mnras}
\bibliography{mnras_template}




\appendix

\section{Very red transient candidates}
\label{sect:appdx:colour_outliers}

As discussed in Sect.~\ref{sect:colour_method_candidates}, we find 84 candidates for glSNe identified based on apparently red colours and relatively bright apparent magnitudes before $r$-band maximum. In all cases, we find no evidence in support of being a glSN and instead all are consistent with non-lensed transients of some kind. We find that each of these candidates can be placed into one of the following five categories: 
\begin{enumerate}
    \item Becoming brighter than the 3$\sigma$ range predicted by our simulations
    \item Red colours due to mis-estimated or poorly-sampled GP fits
    \item Red colours due to declining flux in the $g$-band
    \item Red colours due solely to non-detection limits
    \item Intrinsically red and/or heavily extincted light curves
\end{enumerate}
Here we provide a brief overview of each of these contamination categories and discuss our process of assigning representative candidates to each class. We follow similar processes for other candidates not discussed in detail. A full list of our identified candidates is given in Table~\ref{tab:colour_outliers}, including to which category they belong.

\subsection{Becoming brighter than the 3$\sigma$ range predicted by our simulations}

The brightest peak magnitude predicted by the 3$\sigma$ range from our simulations is $m_r = 14.57$. A small number of transients (3) were selected as outliers after becoming brighter than this limit. In all cases these were very nearby ($\lesssim$35~Mpc), spectroscopically classified SNe~Ia.

\subsection{Red colours due to mis-estimated or poorly-sampled GP fits}

\begin{figure}
\centering
\includegraphics[width=\columnwidth]{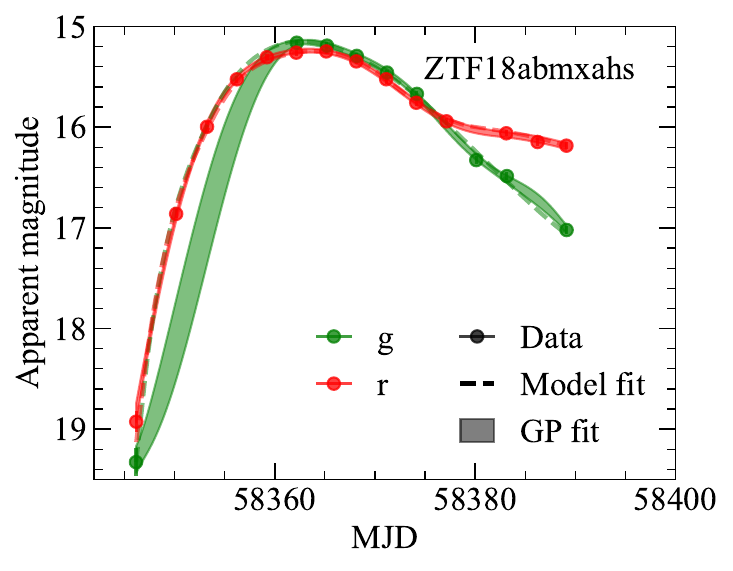}
\caption{Light curve of ZTF18abmxahs, which was identified as a colour outlier in our sample. The ZTF $g$- and $r$-band observations are shown as green and red points, respectively. The 1$\sigma$ ranges in the $g$- and $r$-band light curves predicted by our Gaussian Processes fit are shown as shaded regions. A SALT2 model fit is shown as a dashed line and for which $c = -0.019$. }
\label{fig:colour_method:bad_gp_fit}
\centering
\end{figure}

Based on our SN template fits, we find that 10 transients were flagged due to poorly sampled light curves resulting in the GP fits incorrectly estimating the flux and therefore the colour. An example of this is shown in Fig.~\ref{fig:colour_method:bad_gp_fit} for the case of ZTF18abmxahs, which was spectroscopically classified as a SN~Ia at $z = 0.019$. Figure~\ref{fig:colour_method:bad_gp_fit} shows that ZTF18abmxahs was initially observed shortly after explosion in both the $g$- and $r$-bands, but was not observed again in the $g$-band until around maximum light. Figure~\ref{fig:colour_method:bad_gp_fit} also shows a SALT2 model \citep{guy--07} fit to the light curve of ZTF18abmxahs, which indicates that the poorly sampled $g$-band light curve resulted in our GP fit underestimating the $g$-band flux and hence over-estimating the red colour. Based on our GP fit, we find a $g-r$ colour of $\sim$0.4 -- 1.1 for apparent $r$-band magnitudes of $\sim$15.3 -- 16.8, whereas our SALT2 fit indicates the colour was instead $\sim-0.1$ during the same period. In addition, from our SALT2 fit we find $c = -0.019\pm0.03$ for ZTF18abmxahs, which is typical of normal SNe~Ia. 

\subsection{Red colours due to declining flux in the $g$-band}

\begin{figure}
\centering
\includegraphics[width=\columnwidth]{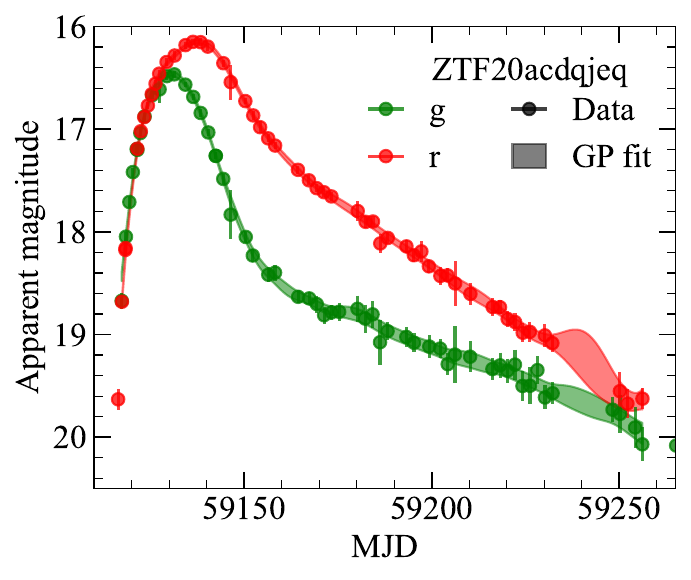}
\caption{As in Fig.~\ref{fig:colour_method:bad_gp_fit} for ZTF20acdqjeq. }
\label{fig:colour_method:g_decline}
\centering
\end{figure}

Within our sample, 6 candidates were identified due to red colours associated with a decline in the $g$-band flux while the $r$-band continued to rise. An example is shown in Fig.~\ref{fig:colour_method:g_decline} for the case of ZTF20acdqjeq, which was spectroscopcially classified as a low-redshift SN~Iax \citep{maguire--23}. ZTF20acdqjeq reached maximum light in the $g$- and $r$-bands on MJD = 59131.0$\pm$0.7 and 59136.9$\pm$0.5, respectively \citep{maguire--23}. The approximately one week gap between the two bands meant that the $r$-band was still within the rising phase as the colour became increasingly red and hence was flagged as an outlier.

\subsection{Red colours due solely to non-detection limits}

\begin{figure}
\centering
\includegraphics[width=\columnwidth]{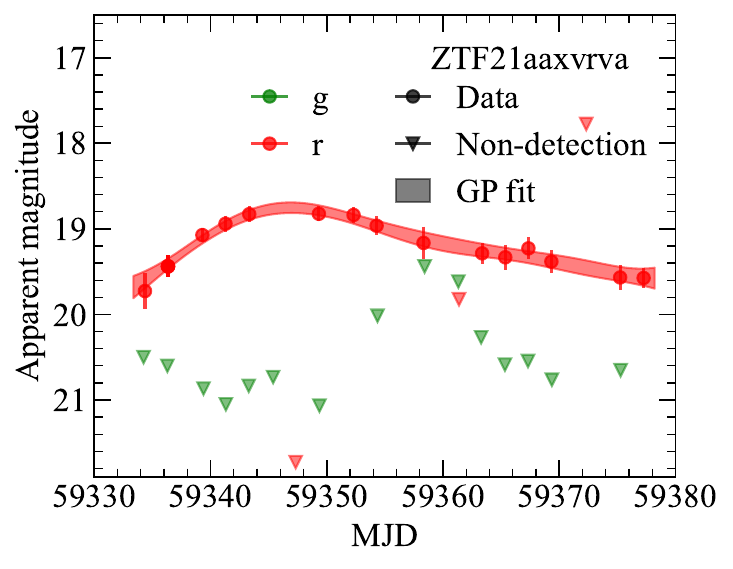}
\caption{As in Fig.~\ref{fig:colour_method:bad_gp_fit} for ZTF21aaxvrva. Non-detections are shown as triangles. }
\label{fig:colour_method:bad_limits}
\centering
\end{figure}

As discussed in Sect.~\ref{sect:colour_method}, we include $g$-band non-detections in our estimates of the colour. Non-detections however do not necessarily indicate an intrinsically red colour and could instead be due to issues with the data itself. An example of this is shown in Fig.~\ref{fig:colour_method:bad_limits} for ZTF21aaxvrva. Despite being spectroscopically classified as a SN~Ia at $z = 0.082$ and having a well-sampled $r$-band light curve, ZTF21aaxvrva was not observed at all in the $g$-band and only non-detection limits exist. If physically real, these $g$-band non-detections imply a $g-r$ colour of $\gtrsim$1.69. Similarly deep non-detection limits also exist in the $r$-band however, including when the SN was relatively bright. For example on MJD = 59343.33 \& 59349.32 the $r$-band magnitudes were $m_r$ = 18.83$\pm$0.10 \& 18.82$\pm$0.09, respectively. A non-detection during the intervening period however, on 59347.32, suggested $m_r \gtrsim 21.73$. We therefore consider these limits in both the $g$- and $r$-band to be related to data issues and not physical. We find 24 other candidates within our sample whereby non-detection limits imply very red colours. In all cases however, following further investigation, we do not consider these to be physically real.

\subsection{Intrinsically red and/or heavily extincted light curves}

Finally, the largest category of outliers identified by our selection criteria are those with intrinsically red colours or heavily extincted light curves. Using the redshift of the likely host galaxy, we use template fitting to assess whether the observed light curve and peak magnitudes are consistent with being an unlensed transient. Here we discuss two representative candidates belonging to this category, one of which was spectroscopically classified while the other was not. 

\par

\begin{figure}
\centering
\includegraphics[width=\columnwidth]{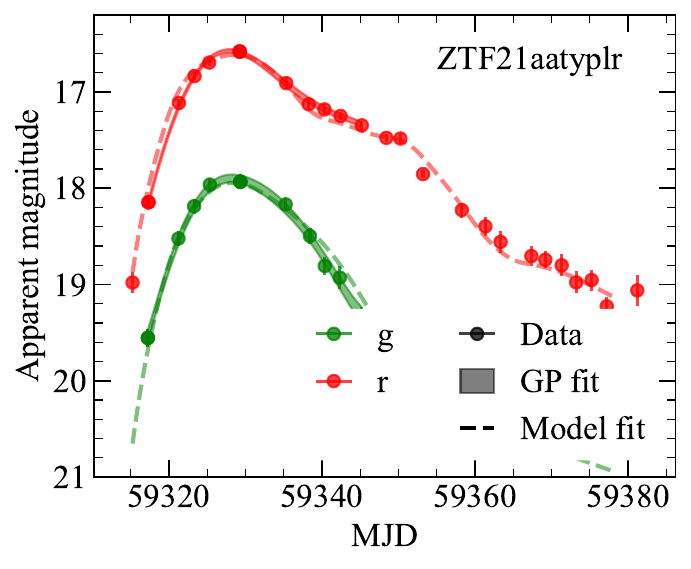}
\caption{As in Fig.~\ref{fig:colour_method:bad_gp_fit} for ZTF21aatyplr, which is shown compared to a SALT2 model fit with $c = 1.26$. }
\label{fig:colour_method:ZTF21aatyplr}
\centering
\end{figure}

ZTF21aatyplr (Fig.~\ref{fig:colour_method:ZTF21aatyplr}) was spectroscopically classified as a SN~Ia \citep{ZTF21aatyplr--class} and reached a peak apparent magnitude of $m_r \sim 16.5$, but showed a relatively flat $g-r$ colour of $\sim$1.35 during its rise. Hence it was flagged as outside our colour limit shown in Fig.~\ref{fig:colour_method}, despite being classified as a SN~Ia. The nearby host galaxy was also observed spectroscopically by \cite{springob--05}, finding a redshift of $z = 0.008$. We find the light curve is in good agreement with a SALT2 model fit and $x_1 = -1.84\pm0.13$ and $c = 1.26\pm0.04$. Publicly available spectra on the TNS also show evidence for strong Na~I~D absorption, indicating heavy dust extinction \citep{poznanski--12}. From our SALT2 fits, and correcting for distance modulus and Milky Way extinction only, we find a peak absolute magnitude of $M_B \sim -14.48$, which is significantly fainter than expected for normal SNe~Ia. Despite showing a large $c$ value, ZTF21aatyplr is well-fit by the SALT2 model and gives a standardised luminosity (following the Tripp equation with $\alpha = 0.148$ and $\beta = 3.112$; \citealt{scolnic--22}) generally within the range expected for SNe~Ia although towards the faint end of the distribution, $M_B \sim -18.66$. Therefore we find no evidence to suggest that ZTF21aatyplr has been strongly lensed and instead is most likely a red, but otherwise relatively normal SNe~Ia.

\par

\begin{figure}
\centering
\includegraphics[width=\columnwidth]{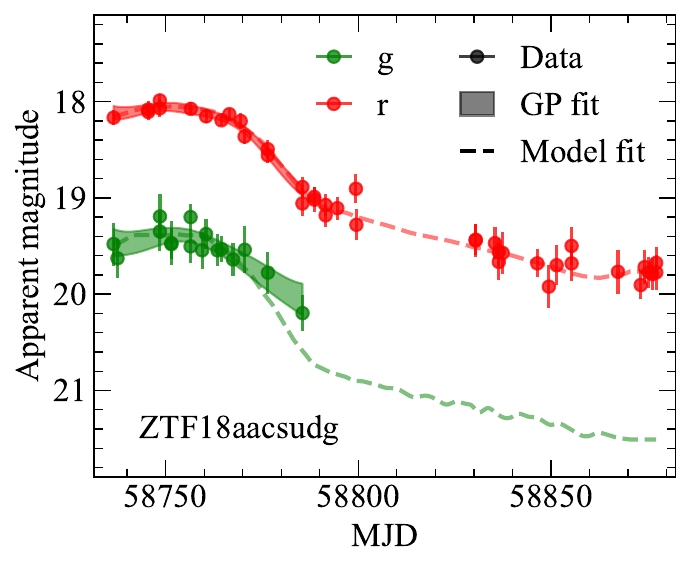}
\caption{As in Fig.~\ref{fig:colour_method:bad_gp_fit} for ZTF18aacsudg, which is shown compared to a template fit of SN~1987A with $A_V = 0.56$.}
\label{fig:colour_method:ZTF18aacsudg}
\centering
\end{figure}

ZTF18aacsudg was not spectroscopically classified, but shows a $g-r$ colour of $\sim$1.34 around peak, $m_r \sim 18.00$. A nearby galaxy (offset by only 0.35\farcs{}) observed spectroscopically by SDSS provides a redshift of $z = 0.025$ \citep{sdss--dr13}. Assuming this redshift and using templates from \cite{vincenzi--19}, we find ZTF18aacsudg to be consistent with the light curve of SN~1987A and a host extinction of A$_{{V}} \sim 0.56$ (Fig.~\ref{fig:colour_method:ZTF18aacsudg}). Even correcting for this extinction ZTF18aacsudg remains an outlier relative to our simulations, but we note that PLAsTiCC did not include any 87A-like templates specifically and therefore our simulations also do not include 87A-like SNe. Assuming a host extinction of A$_{{V}} = 0.56$ and a peak apparent magnitude of $m_g = 19.30$ we find a peak absolute magnitude of $M_g = -15.89$, which is comparable to other 87A-like SNe \citep{taddia--12}.

\clearpage
\onecolumn

\begin{center}
\footnotesize{
\begin{longtable}{lcllccrcc}
\caption{Final candidate glSNe selected by the colour-based method.} \label{tab:colour_outliers} \\
\hline
\multicolumn{1}{c}{\textbf{ZTF name}}	 &	\multicolumn{1}{c}{\textbf{3$\sigma$}} &  \multicolumn{1}{c}{\textbf{Classification}$^{a}$}	& \multicolumn{1}{l}{\textbf{Redshift}$^{b}$}	&	\multicolumn{3}{c}{\textbf{First flagged detection}$^{c}$} &	\multicolumn{1}{c}{\textbf{Outlier}} & \multicolumn{1}{c}{\textbf{Ref.}} \\ 
\multicolumn{1}{c}{}	 &	\multicolumn{1}{c}{\textbf{outlier}} &  \multicolumn{1}{c}{}	& \multicolumn{1}{l}{}	&	\multicolumn{1}{c}{\textbf{MJD}} &  \multicolumn{1}{c}{\textbf{$r$-band magnitude}}	& \multicolumn{1}{c}{\textbf{$g-r$ colour}}	&	\multicolumn{1}{c}{\textbf{category}} & \multicolumn{1}{c}{} \\ 
\hline\hline
\endfirsthead

\multicolumn{9}{c}%
{{\tablename\ \thetable{} -- continued from previous page}} \\
\hline
\multicolumn{1}{c}{\textbf{ZTF name}}	 &	\multicolumn{1}{c}{\textbf{3$\sigma$}} &  \multicolumn{1}{c}{\textbf{Classification}$^{a}$}	& \multicolumn{1}{l}{\textbf{Redshift}$^{b}$}	&	\multicolumn{3}{c}{\textbf{First flagged detection}$^{c}$} &	\multicolumn{1}{c}{\textbf{Outlier}} & \multicolumn{1}{c}{\textbf{Ref.}} \\ 
\multicolumn{1}{c}{}	 &	\multicolumn{1}{c}{\textbf{outlier}} &  \multicolumn{1}{c}{}	& \multicolumn{1}{l}{}	&	\multicolumn{1}{c}{\textbf{MJD}} &  \multicolumn{1}{c}{\textbf{$r$-band magnitude}}	& \multicolumn{1}{c}{\textbf{$g-r$ colour}}	&	\multicolumn{1}{c}{\textbf{category}} & \multicolumn{1}{c}{} \\ 
\hline\hline
\endhead

\hline \multicolumn{9}{|r|}{{Continued on next page}} \\ \hline
\endfoot

\endlastfoot
ZTF18aacsudg  &  N   &  \textit{SN 1987A-like} 	&	0.025$^{\textrm{†}}$	&  58745.49 & 18.03$\pm$0.04 & 1.34$\pm$0.11 			& 	v.	&		\\
ZTF18aaxxyhp  &  Y   &  \textit{Nuclear/Ibc?} 	&	0.022$^{\textrm{†}}$	&  58278.33 & 18.73$\pm$0.08 & $\gtrsim$ 2.12$\pm$0.08 	& 	iv.	&		\\
ZTF18abgmcmv  &  Y   &  Ia-91T 					&	0.019					&  58323.26 & 16.59$\pm$0.03 & 0.48$\pm$0.04 			& 	v.	&		\\
ZTF18abmxahs  &  N   &  Ia  					&	0.015					&  58350.17 & 16.85$\pm$0.04 & 1.13$\pm$0.40 			& 	ii.	&		\\
ZTF18ackaxnw  &  N   &  \textit{Ia-91bg/Ibc?} 	&	0.021$^{\textrm{†}}$	&  58435.27 & 18.85$\pm$0.10 & $\gtrsim$1.32$\pm$0.10 	& 	iv.	&		\\
ZTF19aadttht  &  Y   &  Ic 						&	0.006					&  58509.49 & 15.97$\pm$0.03 & 0.63$\pm$0.07 			& 	v.	&		\\
ZTF19aadyppr  &  Y   &  ILRT 					&	0.002					&  58522.49 & 16.81$\pm$0.07 & 0.89$\pm$0.08 			& 	v.	&		\\
ZTF19aafncsv  &  N   &  Ia 						&	0.037					&  58526.54 & 17.13$\pm$0.05 & 1.23$\pm$0.51 			& 	ii.	&		\\
ZTF19aanyuyh  &  Y   &  Ia 						&	0.025$^{\textrm{†}}$	&  59051.33 & 17.39$\pm$0.02 & 0.73$\pm$0.04 			& 	v.	&		\\
ZTF19aarnqzw  &  N   &  Ia 						&	0.028					&  58606.34 & 16.33$\pm$0.12 & 0.50$\pm$0.26 			& 	ii.	&		\\
ZTF19aavkvpw  &  N   &  Ic-BL 					&	0.027					&  58642.17 & 17.46$\pm$0.03 & 0.87$\pm$0.07 			& 	iii.	&		\\
ZTF19abeihhp  &  N   &  \textit{Nuclear/Ia?} 	&	0.157$\pm$0.063			&  58672.20 & 19.14$\pm$0.10 & 1.47$\pm$0.21 			& 	ii.	&		\\
ZTF19abguibf  &  Y   &  \textit{Mira variable} 	&	0.000$^{\textrm{†}}$	&  58706.47 & 19.12$\pm$0.04 & 1.40$\pm$0.09 			& 	v.	&		\\
ZTF19abqgyzx  &  N   &  \textit{Stellar?} 		&	0.000$^{\textrm{†}}$	&  58716.49 & 15.97$\pm$0.03 & 0.99$\pm$0.95 			& 	ii.	&		\\
ZTF19abrnpst  &  Y   &  \textit{II} 			&	0.081$\pm$0.038			&  58723.39 & 19.12$\pm$0.05 & $\gtrsim$1.55$\pm$0.05 	& 	iv.	&		\\
ZTF19abucwzt  &  N   &  Ib 						&	0.017					&  58909.11 & 18.99$\pm$0.03 & 1.27$\pm$0.07 			& 	v.	&	 1  \\
ZTF19abxqppy  &  N   &  IIb 					&	0.014					&  58737.13 & 18.09$\pm$0.07 & 1.16$\pm$0.18 			& 	v.	&		\\
ZTF19achaspq  &  N   &  Ia 						&	0.016					&  58790.10 & 16.26$\pm$0.38 & 0.52$\pm$0.39 			& 	v.	&		\\
ZTF19acihlft  &  N   &  Ia 						&	0.020					&  58794.15 & 14.89$\pm$0.44 & 0.09$\pm$0.45 			& 	ii.	&		\\
ZTF19acmxidf  &  Y   &  Ia 						&	0.014					&  58799.46 & 16.72$\pm$0.04 & 1.04$\pm$0.08 			& 	v.	&		\\
ZTF19acnzkph  &  Y   &  Ia 						&	0.018					&  58798.17 & 16.85$\pm$0.04 & 0.76$\pm$0.07 			& 	v.	&		\\
ZTF19acxxwvi  &  N   &  IIb 					&	0.011					&  58831.55 & 17.15$\pm$0.04 & 0.70$\pm$0.07 			& 	v.	&		\\
ZTF19adakmbh  &  N   &  II 						&	0.018					&  58863.11 & 18.41$\pm$0.09 & 1.14$\pm$0.21 			& 	v.	&		\\
ZTF19adaxzax  &  N   &  II 						&	0.030					&  58855.19 & 16.00$\pm$0.04 & 0.47$\pm$0.12 			& 	v.	&		\\
ZTF20aaelulu  &  Y   &  Ic 						&	0.005					&  58859.48 & 14.74$\pm$0.03 & 0.11$\pm$0.04 			& 	v.	&		\\
ZTF20aamifit  &  Y   &  Ia 						&	0.060					&  58886.56 & 17.62$\pm$0.06 & $\gtrsim$1.55$\pm$0.06 	& 	iv.	&		\\
ZTF20aaocqkr  &  Y   &  IIn 					&	0.024					&  58898.36 & 17.72$\pm$0.04 & $\gtrsim$3.29$\pm$0.04 	& 	iv.	&		\\
ZTF20aattotq  &  Y   &  Ia 						&	0.014					&  58936.44 & 16.20$\pm$0.05 & 0.36$\pm$0.06 			& 	v.	&		\\
ZTF20aaxbvkt  &  N   &  Ia 						&	0.040					&  58973.31 & 18.53$\pm$0.08 & 1.17$\pm$0.57 			& 	ii.	&		\\
ZTF20aayqjpv  &  Y   &  Ia 						&	0.032					&  58973.33 & 17.93$\pm$0.13 & $\gtrsim$1.60$\pm$0.13 	& 	iv.	&		\\
ZTF20abasxmb  &  N   &  Ia 						&	0.025					&  58995.25 & 16.68$\pm$0.12 & 0.50$\pm$0.19 			& 	ii.	&		\\
ZTF20abchbds  &  N   &  Ia 						&	0.011					&  58999.36 & 17.03$\pm$0.04 & 0.55$\pm$0.14 			& 	ii.	&		\\
ZTF20abefbpl  &  Y   &  Ic 						&	0.042					&  59018.21 & 18.32$\pm$0.04 & 1.16$\pm$0.09 			& 	v.	&		\\
ZTF20abhjwvh  &  N   &  II 						&	0.010					&  59037.19 & 14.92$\pm$0.02 & 0.21$\pm$0.58 			& 	v.	&		\\
ZTF20abhmqdn  &  Y   &  Ia 						&	0.038					&  59027.28 & 18.91$\pm$0.04 & $\gtrsim$1.63$\pm$0.04 	& 	iv.	&		\\
ZTF20abpmqnr  &  Y   &  IIn 					&	0.022					&  59063.42 & 17.14$\pm$0.02 & 0.66$\pm$0.05 			& 	v.	&		\\
ZTF20abuovvw  &  N   &  \textit{II/Ibc?} 		&	0.087$\pm$0.040			&  59110.28 & 19.13$\pm$0.04 & 1.38$\pm$0.12 			& 	iii.	&		\\
ZTF20abynbhh  &  Y   &  \textit{Nuclear/I?} 	&	0.072$^{\textrm{†}}$	&  59312.17 & 18.81$\pm$0.05 & $\gtrsim$1.34$\pm$0.05 	& 	iv.	&		\\
ZTF20acdqjeq  &  Y   &  Iax 					&	0.017					&  59134.28 & 16.17$\pm$0.02 & 0.29$\pm$0.03 			& 	iii.	&     2 \\
ZTF20acduffd  &  Y   &  Ia 						&	0.020					&  59116.40 & 18.95$\pm$0.07 & $\gtrsim$1.67$\pm$0.07 	& 	iv.	&		\\
ZTF20acimjmc  &  Y   &  Ia 						&	0.033					&  59136.49 & 18.94$\pm$0.07 & $\gtrsim$1.61$\pm$0.07 	& 	iv.	&		\\
ZTF20aclwclm  &  N   &  Ia 						&	0.008					&  59154.16 & 14.30$\pm$0.09 & 0.33$\pm$0.18 			& 	i.	&		\\
ZTF20acpjqxp  &  Y   &  Ib 						&	0.007					&  59173.22 & 15.36$\pm$0.01 & 0.08$\pm$0.02 			& 	v.	&		\\
ZTF20acrzwvx  &  N   &  II 						&	0.010					&  59186.43 & 16.58$\pm$0.08 & 0.39$\pm$0.09 			& 	v.	&		\\
ZTF20actxqkp  &  Y   &  \textit{Stellar?} 		&	0.000$^{\textrm{†}}$	&  59202.13 & 16.98$\pm$0.10 & 1.06$\pm$0.25 			& 	v.	&		\\
ZTF20acungqk  &  Y   &  \textit{Nuclear/Ibc?} 	&	0.043$\pm$0.013			&  59193.23 & 18.63$\pm$0.05 & 1.24$\pm$0.20 			& 	iii.	&		\\
ZTF20acynjjo  &  Y   &  Ia 						&	0.015					&  59217.10 & 15.72$\pm$0.03 & 0.24$\pm$0.04 			& 	v.	&		\\
ZTF21aaabwfu  &  N   &  IIb 					&	0.011					&  59229.53 & 18.21$\pm$0.05 & 1.12$\pm$0.11 			& 	v.	&		\\
ZTF21aaaubig  &  Y   &  Ic 						&	0.009					&  59224.31 & 16.69$\pm$0.02 & 0.48$\pm$0.03 			& 	v.	&		\\
ZTF21aacuera  &  Y   &  \textit{Ia} 			&	0.084$\pm$0.029			&  59227.55 & 19.06$\pm$0.09 & $\gtrsim$1.37$\pm$0.09 	& 	iv.	&		\\
ZTF21aafnunh  &  N   &  Ic-BL 					&	0.030					&  59250.19 & 17.02$\pm$0.04 & 0.62$\pm$0.79 			& 	ii.	&		\\
ZTF21aagtvjq  &  Y   &  II 						&	0.083					&  59253.51 & 18.66$\pm$0.04 & $\gtrsim$2.02$\pm$0.04 	& 	iv.	&		\\
ZTF21aamwqim  &  N   &  II 						&	0.026					&  59280.29 & 18.33$\pm$0.03 & 1.20$\pm$0.48 			& 	iii.	&		\\
ZTF21aaqytjr  &  N   &  Ia 						&	0.003					&  59309.28 & 14.44$\pm$0.02 & -0.06$\pm$0.07 			& 	i.	&		\\
ZTF21aasaxfg  &  Y   &  Ia 						&	0.070					&  59311.47 & 19.24$\pm$0.06 & $\gtrsim$1.56$\pm$0.06 	& 	iv.	&		\\
ZTF21aatyplr  &  Y   &  Ia 						&	0.008					&  59317.25 & 18.13$\pm$0.04 & 1.35$\pm$0.07 			& 	v.	&		\\
ZTF21aavheiv  &  Y   &  Ia 						&	0.053					&  59322.26 & 18.93$\pm$0.04 & $\gtrsim$1.67$\pm$0.04 	& 	iv.	&		\\
ZTF21aavodst  &  Y   &  Ia 						&	0.019					&  59323.16 & 17.10$\pm$0.04 & 0.81$\pm$0.07 			& 	v.	&		\\
ZTF21aaxtctv  &  N   &  Ic 						&	0.014					&  59350.19 & 16.69$\pm$0.02 & 0.48$\pm$0.04 			& 	iii.	&		\\
ZTF21aaxvkae  &  N   &  \textit{Ia} 			&	0.142$\pm$0.063			&  59341.28 & 19.32$\pm$0.06 & $\gtrsim$1.40$\pm$0.06 	& 	iv.	&		\\
ZTF21aaxvrva  &  Y   &  Ia 						&	0.082					&  59339.30 & 19.07$\pm$0.05 & $\gtrsim$1.69$\pm$0.05 	& 	iv.	&		\\
ZTF21aaydxoo  &  Y   &  IIn 					&	0.022					&  59346.30 & 16.87$\pm$0.02 & 0.56$\pm$0.03 			& 	v.	&		\\
ZTF21aazbico  &  Y   &  Ia 						&	0.083					&  59344.38 & 19.17$\pm$0.05 & $\gtrsim$1.68$\pm$0.05 	& 	iv.	&		\\
ZTF21abbyfjc  &  Y   &  \textit{Ia/Ibc?} 		&	0.047$^{\textrm{†}}$	&  59362.25 & 18.18$\pm$0.06 & $\gtrsim$1.58$\pm$0.06 	& 	iv.	&		\\
ZTF21abcgaln  &  N   &  Ic 						&	0.010					&  59364.35 & 16.13$\pm$0.03 & 0.28$\pm$0.04 			& 	v.	&		\\
ZTF21abckuxr  &  Y   &  \textit{IIb?} 			&	0.072$\pm$0.028			&  59359.42 & 17.91$\pm$0.18 & $\gtrsim$1.43$\pm$0.18 	& 	iv.	&		\\
ZTF21abcxswe  &  Y   &  Ia 						&	0.032					&  59366.35 & 18.70$\pm$0.07 & $\gtrsim$1.71$\pm$0.07 	& 	iv.	&		\\
ZTF21abfaohe  &  N   &  Ia 						&	0.009					&  59384.20 & 14.65$\pm$0.03 & $-$0.08$\pm$0.04 	    & 	v.	&		\\
ZTF21abfmbix  &  N   &  Ia 						&	0.009					&  59386.19 & 15.08$\pm$0.26 & 0.26$\pm$0.27 			& 	v.	&		\\
ZTF21abiuvdk  &  Y   &  Ia 						&	0.004					&  59400.41 & 13.80$\pm$0.02 & 0.26$\pm$0.03 			& 	i.	&	3   \\
ZTF21abjcliz  &  Y   &  II 						&	0.031					&  59447.21 & 18.95$\pm$0.04 & 1.32$\pm$0.21 			& 	v.	&		\\
ZTF21abjyiiw  &  Y   &  IIb 					&	0.005					&  59406.38 & 17.74$\pm$0.04 & 0.65$\pm$0.06 			& 	v.	&		\\
ZTF21abvbdyr  &  N   &  II 						&	0.020					&  59463.21 & 17.21$\pm$0.02 & 0.68$\pm$0.04 			& 	v.	&		\\
ZTF21abxnuwt  &  Y   &  \textit{Nuclear/I?} 	&	0.016$^{\textrm{†}}$	&  59465.49 & 18.48$\pm$0.11 & $\gtrsim$1.18$\pm$0.11 	& 	iv.	&		\\
ZTF21acckcni  &  Y   &  IIn 					&	0.043					&  59472.49 & 18.67$\pm$0.10 & $\gtrsim$1.22$\pm$0.10 	& 	iv.	&		\\
ZTF21acdontl  &  Y   &  II 						&	0.010					&  59500.46 & 16.86$\pm$0.03 & 0.87$\pm$0.07 			& 	v.	&		\\
ZTF21aceehxt  &  Y   &  Ib 						&	0.013					&  59496.37 & 17.48$\pm$0.03 & 0.97$\pm$0.07 			& 	v.	&		\\
ZTF21acenkuf  &  N   &  Ia 						&	0.012					&  59497.34 & 17.10$\pm$0.10 & 0.91$\pm$0.11 			& 	v.	&		\\
ZTF21achkqhi  &  Y   &  II 						&	0.031					&  59503.36 & 18.33$\pm$0.06 & $\gtrsim$1.18$\pm$0.06 	& 	iv.	&		\\
ZTF21acizlsw  &  Y   &  \textit{I} 				&	0.013$\pm$0.008			&  59518.33 & 18.90$\pm$0.05 & $\gtrsim$1.48$\pm$0.05 	& 	iv.	&		\\
ZTF21acjoorl  &  N   &  Ic				 		&	0.010					&  59526.45 & 18.04$\pm$0.03 & 0.66$\pm$0.07 			& 	v.	&		\\
ZTF21aclyyfm  &  Y   &  II  					&	0.005					&  59530.49 & 16.98$\pm$0.08 & 0.71$\pm$0.10 			& 	v.	&		\\
ZTF21acpmcgo  &  N   &  Ia 						&	0.016					&  59538.34 & 17.31$\pm$0.07 & 0.31$\pm$0.42 			& 	v.	&		\\
ZTF22aaaxkar  &  N   &  Ia 						&	0.020					&  59639.52 & 17.06$\pm$0.07 & 0.25$\pm$0.10 			& 	v.	&		\\
\hline\hline
\multicolumn{9}{p{16cm}}{$^{a}$ Classifications given in italics are based on photometric templates or cross-matching. All other classifications are based on spectroscopic observations.} \\
\multicolumn{9}{p{16cm}}{$^{b}$ PS1 photometric redshifts are given with the associated uncertainty, while $^{\textrm{†}}$ denotes redshifts of the nearby, likely host galaxy. All other redshifts are based on spectroscopic observations of the transient.} \\
\multicolumn{9}{p{16cm}}{$^{c}$ Magnitudes and colours are given corrected for Milky Way extinction only.} \\
\multicolumn{9}{p{16cm}}{References: (1) \cite{sollerman--20}, (2) \cite{maguire--23}, (3) \cite{dhawan--22b}.} \\
\end{longtable}
}
\end{center}

\clearpage
\twocolumn


\section{Light curve inconsistent with SN~Ia at photometric redshift of nearby elliptical galaxy candidates}
\label{sect:appdx:model_outliers}

Following the implementation described by \cite{goldstein--18--lens}, we find 29 candidates for glSNe identified on the basis of significant outliers relative to a SALT2 model fit at the redshift of the nearby elliptical host galaxy. In all cases, we find no evidence in support of glSNe and instead all are consistent with various types of non-lensed transients. Each of the candidates can be placed into one of the following five categories: 

\begin{enumerate}
    \item Normal SNe~Ia
    \item Peculiar SNe~Ia
    \item Other transients
    \item Light curves with multiple peaks
    \item Incorrect host redshifts
\end{enumerate}

In the following section we briefly describe our process of determining the reason why representative candidates have been identified and assigning them to these categories. For candidates not discussed in detail here we follow a similar process. A full list of our candidates selected using this method is given in Table~\ref{tab:model_outliers}, including to which category they belong.

\subsection{Normal SNe~Ia}

\begin{figure}
\centering
\includegraphics[width=\columnwidth]{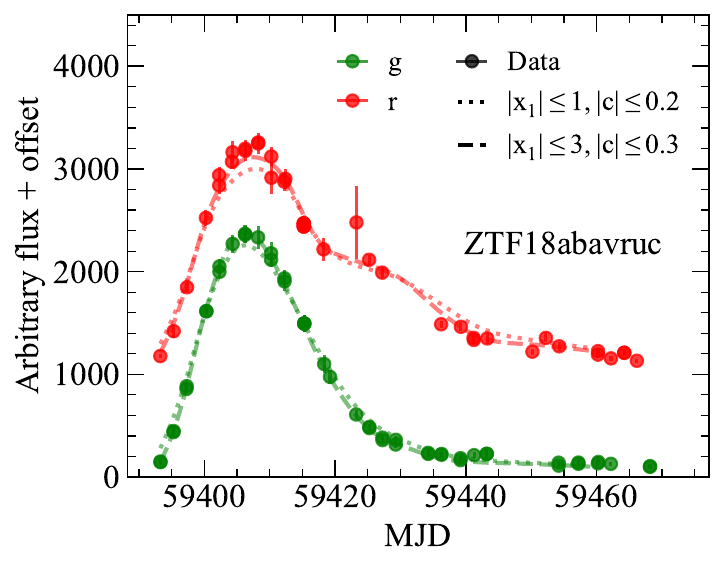}
\caption{Light curve of ZTF18abavruc compared to SALT2 template fits assuming either $|x_1| \leq 1, |c| \leq 0.2$ or $|x_1| \leq 3, |c| \leq 0.3$. }
\label{fig:outlier_method:normal_ia}
\centering
\end{figure}

A number of candidates identified using this method were spectroscopically classified as SNe~Ia or show light curves consistent with being normal SNe~Ia, but also show at least one outlier relative to the SALT2 model fit. In some cases these outliers were due to spurious detections or fluctuations in the light curve. For the rest however, we find their light curves can be well fit by extending the SALT2 boundaries suggested by \cite{goldstein--18--lens} from $|x_1| \leq 1$ and $|c| \leq 0.2$ to $|x_1| \leq 3$ and $|c| \leq 0.3$. These extended boundaries are still within the limits of typical cosmological samples (e.g. \citealt{dhawan--22}). An example is shown in Fig.~\ref{fig:outlier_method:normal_ia} for the case of ZTF18abavruc. Figure~\ref{fig:outlier_method:normal_ia} shows that the restricted $x_1$ and $c$ fitting boundaries are not able to reproduce the observed light curve shape, particularly up to and including maximum light. The extended SALT2 boundaries however provide much better agreement with the overall shape of the light curve and do not produce any 5$\sigma$ outliers.

\subsection{Peculiar SNe~Ia}

\begin{figure}
\centering
\includegraphics[width=\columnwidth]{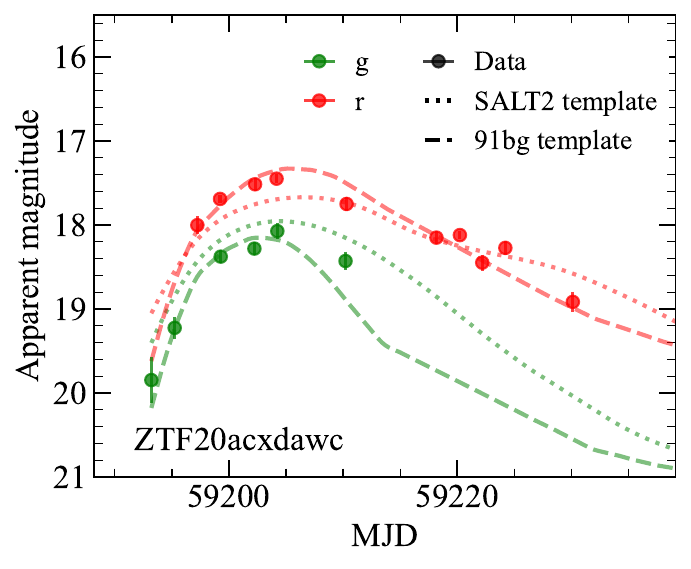}
\caption{Light curve of ZTF20acxdawc compared to SALT2 and SN~1991bg-like templates. }
\label{fig:outlier_method:91bg}
\centering
\end{figure}

Three of our candidates were spectroscopically classified as SNe~Ia, but show significant deviations from the SALT2 model, even with extended boundaries for $x_1$ and $c$. For these SNe, we fit their light curves using sub-luminous, SN~1991bg-like templates and find they are consistent. Figure~\ref{fig:outlier_method:91bg} shows an example case of ZTF20acxdawc, along with the SALT2 and 91bg-like template fits.

\subsection{Other transients}

The most numerous source of contamination found by our selection criteria was other types of transients that are not normal SNe~Ia. As in Sect.~\ref{sect:colour_method_candidates}, we fit the light curves using templates of various types of SNe and assume the redshift of the elliptical galaxy. For approximately four candidates we are unable to find a suitable match with existing templates, but based on the observed light curve we speculate that these are most likely AGN that have passed our initial cuts.

\par

\begin{figure}
\centering
\includegraphics[width=\columnwidth]{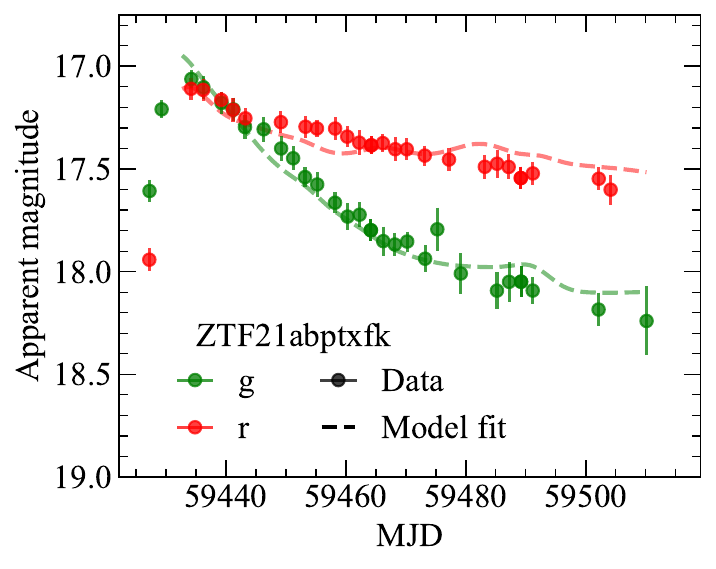}
\caption{Light curve of ZTF21abptxfk compared to the SN~II SN~2005gi. }
\label{fig:outlier_method:ZTF21abptxfk}
\centering
\end{figure}

Three of our selected candidates are consistent with SNe~II light curves. While none of these SNe show obvious signs of spiral host galaxies in archival imaging, the uncertainties on the host galaxy colours mean they are also consistent with falling outside our selection cuts and therefore may not truly be ellipticals. ZTF21abptxfk was spectroscopically classified as a SN~II however and has host galaxy colours that fall well within our selection criteria. The light curve shows strong similarities to SNe~II, which is consistent with the spectroscopic classification. In Fig.~\ref{fig:outlier_method:ZTF21abptxfk} we show a comparison against the template of the SN~II SN~2005gi \citep{kessler--10}. While no clearly identifiable spiral arms are visible in either PS1 or SDSS archival imaging, we note that the host is classified as a star in the SDSS catalogue. Given the similarities to SNe~II light curves, ZTF21abptxfk may be one of the few SNe~II hosted in elliptical galaxies \citep{irani--22}. 

\subsection{Light curves with multiple peaks}

Two of our identified candidates, ZTF18abdgwvs and ZTF19aawhagd, show evidence of multiple peaks, which could be indicative of sibling pairs (i.e. multiple SNe in the same galaxy; \citealt{graham--22, kelsey--23}) or additional AGN activity. We note however that the photometric uncertainties for observations of both objects are relatively large, making a definite identification difficult. 

\par

\begin{figure*}
\begin{tabular}{cc}
  \includegraphics[width=\columnwidth]{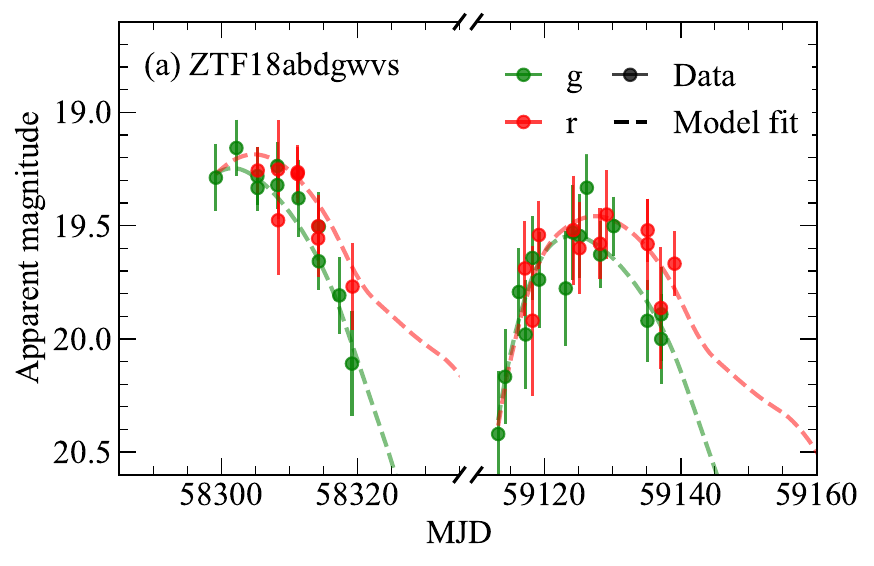} &   
  \includegraphics[width=\columnwidth]{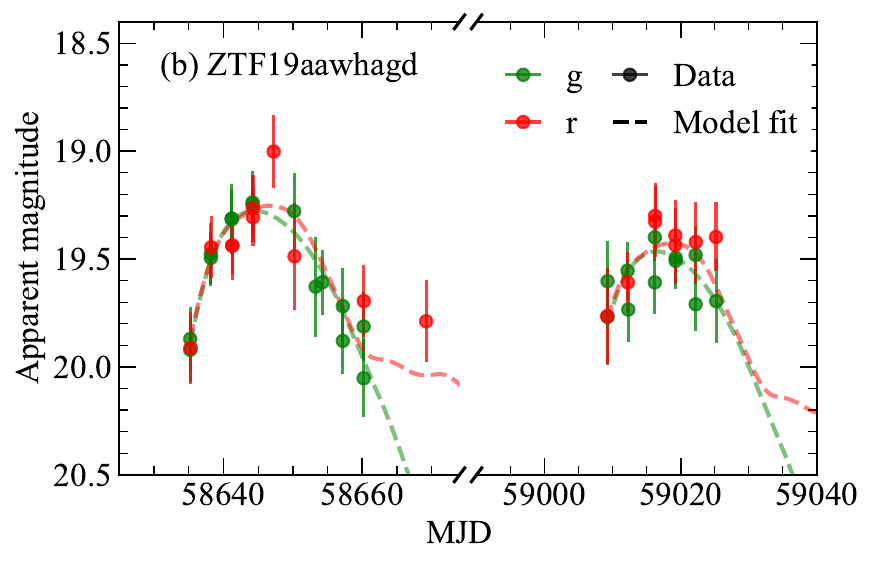}
\end{tabular}
\caption{\textit{Panel a: }Light curve of ZTF18abdgwvs. Two peaks, separated by more than two years, are clearly apparent, causing it to be flagged as an outlier. Dashed lines show SALT2 fits to each individual light curve.
\textit{Panel b: }Light curve of ZTF19aawhagd. As in \textit{Panel a}.}
\label{fig:outlier_method:siblings?}
\end{figure*}

In the case of ZTF18abdgwvs, we find a photometric redshift for the elliptical of $z = 0.122\pm0.027$. Fitting each light curve with the SALT2 model at this redshift, we find both are consistent with SNe~Ia although the uncertainty on fitted parameters are relatively large (Fig.~\ref{fig:outlier_method:siblings?}(a)). The first light curve peaks at MJD = 58\,303.46$\pm$1.65 with $x_1 = 1.74\pm0.91$ and $c = 0.03\pm0.02$, while the second peaks at MJD = 59\,126.12$\pm$0.70 with $x_1 = 1.92\pm0.71$ and $c = 0.05\pm0.06$. While we cannot specifically rule out AGN activity, the median coordinates for both light curves are also separated by $\sim$0.89\farcs{}, indicating these may indeed be two separate events. We also note that given the large uncertainties, both light curves are generally consistent with having the same SALT2 parameters and therefore are qualitatively consistent with expectations for glSNe -- the same SN reappearing with some time delay. As both light curves can be fit with the likely host galaxy redshift and show relatively blue colours, this is unlikely to be the case. In addition, such a significant time delay (more than 2 years) would likely only result from very high-mass lensing systems and archival imaging shows no indication of any lensing features, which may be expected in this case.

\par

Our final candidate, ZTF19aawhagd also shows evidence of multiple light curve peaks (Fig.~\ref{fig:outlier_method:siblings?}(b)). NED provides a spectroscopic redshift of $z = 0.0116$ for the nearby galaxy. The first light curve is consistent with SALT2 at this redshift and peak MJD = $58\,645.02\pm0.53$, $x_1 = 1.33\pm0.72$, and $c =  0.14\pm0.07$. For the second light curve, we find best-fitting SALT2 parameters of peak MJD = $59\,017.01\pm0.96$, $x_1 = 1.25\pm1.05$, and $c = 0.16\pm0.04$. In this case, the median coordinates of the two light curves are separated by only $\sim$0.19\farcs{}, indicating it could be one event, but the $\sim$1\farcs{} from the host galaxy centre makes AGN activity somewhat more unlikely. Again, the large uncertainties for best-fitting parameters mean that ZTF19aawhagd is also generally consistent with expectations for the same SN reappearing with a $\sim$1 year time-delay. As with ZTF18abdgwvs however, based on the colour and long delay time required we find a glSN origin to be unlikely. We note that \cite{graham--22} searched for sibling pairs within the ZTF Bright Transient Survey (BTS; \citealt{fremling--20}). Neither ZTF18abdgwvs nor ZTF19aawhagd reached the required threshold for automatic spectroscopic classification and hence were not included in BTS, however their identification shows that ZTF may contain additional previously unidentified sibling pairs due to being categorised as a single transient.

\subsection{Incorrect host redshifts}

Finally, two candidates were flagged due to fits with likely incorrect host redshifts. ZTF18abcsgvj was spectroscopically classified as a SN~Ia at a redshift of $z = 0.06$, while our estimate of the photometric redshift from PS1 is $z = 0.17\pm0.02$. The photometric redshift from SDSS ($z = 0.07\pm0.01$) however is consistent with the spectroscopic redshift and we find the light curve is well matched by SALT2 with the correct redshift. Assuming the redshift of the nearby elliptical for ZTF21achknpr ($z = 0.29\pm0.03$) places it outside the luminosity range expected for normal SNe~Ia. ZTF21achknpr is offset from this galaxy by 2.99\farcs{}, however for another galaxy at 3.14\farcs{} we find a photometric redshift of $z = 0.10\pm0.05$, which provides a good match to the light curve assuming a SN~Ia.

\begin{table*}
\begin{center}
\caption{Final candidate glSNe selected by the outlier-based method.}
\label{tab:model_outliers}
\begin{tabular}{llllllc}
\hline
\textbf{ZTF name}  	&	\textbf{Classification}$^{a}$ 	&                          &    \textbf{Elliptical galaxy}$^{b}$ 	&                       &                     & Outlier  \\
                	&                                  	& \textbf{Photometric $z$} &	\textbf{Separation}                 &	\textbf{$W2 - W3$} 	& \textbf{$NUV - r$}. & category \\
\hline
\hline
ZTF18aabdajx    &    TDE							&    	 0.04$\pm$0.02 & 0.27\farcs{} &  $\textless-$0.57	& 			 		& iii.	 	\\
ZTF18aasvknh    &    \textit{AGN?}					&    	 0.11$\pm$0.03 & 0.11\farcs{} &  \phn{}\phn{}$-$0.03$\pm$0.36 	&  					& iii.	 	\\
ZTF18aavxiih    &    \textit{II?}					&    	 0.12$\pm$0.03 & 1.76\farcs{} &  \phn{}\phn{}~~~0.26$\pm$0.21 		&  					& iii.	 	\\
ZTF18abavruc    &    Ia								&    	 0.05$\pm$0.02 & 0.11\farcs{} &  \phn{}\phn{}$-$0.86$\pm$0.46 	&  					& i.	 	\\
ZTF18abcsgvj    &    Ia								&    	 0.17$\pm$0.02 & 1.88\farcs{} &  \phn{}\phn{}$-$0.99$\pm$0.40 	&  					& v.	 	\\
ZTF18abdgwvs    &    \textit{Sibling SNe/AGN?}		&    	 0.12$\pm$0.03 & 0.40\farcs{} &  \phn{}\phn{}~~~0.25$\pm$0.26 		&  					& iv.	 	\\
ZTF18abhhxcp    &    Ia								&    	 0.07$\pm$0.01 & 1.91\farcs{} &  \phn{}\phn{}$-$1.01$\pm$0.13 	&  					& i.	 	\\
ZTF18abzrsuh    &    \textit{AGN?}					&    	 0.11$\pm$0.04 & 0.33\farcs{} &  \phn{}\phn{}$-$0.08$\pm$0.21 	&  3.01$\pm$0.15 	& iii.	 	\\
ZTF18acbvgqw    &    Ia								&    	 0.01$\pm$0.01 & 1.92\farcs{} &  \phn{}\phn{}$-$1.31$\pm$0.03 	&  5.39$\pm$0.04	& i.	 	\\
ZTF18acqgvrt    &    \textit{Ia}					&    	 0.05$\pm$0.02 & 0.37\farcs{} &  $\textless-$0.94 	&  					& i. 	 	\\
ZTF18acqywlx    &    \textit{AGN?}					&    	 0.13$\pm$0.03 & 0.17\farcs{} &  \phn{}\phn{}~~~0.28$\pm$0.49 		&  					& iii. 		\\
ZTF18acslpba    &    Ia								&    	 0.03$\pm$0.01 & 3.67\farcs{} &  \phn{}\phn{}$-$0.76$\pm$0.05 	&  6.20$\pm$0.12 	& i.	 	\\
ZTF19aawhagd    &    \textit{Sibling SNe/AGN?}		&    	 0.14$\pm$0.05 & 1.22\farcs{} &  \phn{}\phn{}$-$0.23$\pm$0.50 	&  					& iv. 		\\
ZTF19aayvyeo    &    \textit{AGN?}					&    	 0.28$\pm$0.17 & 0.14\farcs{} &  \phn{}\phn{}~~~0.25$\pm$0.05 		&  					& iii. 		\\
ZTF20abfcszi    &    TDE							&    	 0.09$\pm$0.03 & 0.10\farcs{} &  $\textless-$0.14 	&  					& iii. 		\\
ZTF20abhvnzc    &    Ia								&     	 0.03$\pm$0.01 & 3.56\farcs{} &  \phn{}\phn{}$-$1.21$\pm$0.05 	&  5.44$\pm$0.07 	& ii. 		\\
ZTF20ablxtlp    &    Ia								&     	 0.06$\pm$0.01 & 1.38\farcs{} &  \phn{}\phn{}$-$0.92$\pm$0.42 	&  					& ii. 		\\
ZTF20abqbzuv    &    Ia								&     	 0.03$\pm$0.01 & 1.86\farcs{} &  \phn{}\phn{}$-$1.40$\pm$0.11 	&  6.20$\pm$0.21 	& i.		\\
ZTF20abxyims    &    \textit{SLSN?}					&     	 0.21$\pm$0.03 & 0.62\farcs{} &  $\textless$~\phs{}0.30	&  					& ii. 		\\
ZTF20acitpfz    &    TDE							&     	 0.07$\pm$0.03 & 0.19\farcs{} &  $\textless-$0.07	&  4.73$\pm$0.17 	& iii.		\\
ZTF20acwjnux    &    \textit{IIn?}					&    	 0.11$\pm$0.02 & 2.61\farcs{} &  \phn{}\phn{}~~~0.05$\pm$0.32 		&  3.69$\pm$0.17 	& iii.		\\
ZTF20acxdawc    &    Ia								&    	 0.04$\pm$0.01 & 0.89\farcs{} &  \phn{}\phn{}~~~0.08$\pm$0.04 		&  					& ii. 		\\		
ZTF21aaanpyy    &    \textit{II?}					&    	 0.07$\pm$0.03 & 4.53\farcs{} &  \phn{}\phn{}~~~0.39$\pm$0.12 		&  3.70$\pm$0.15 	& iii.		\\
ZTF21abcgnqn    &    TDE							&    	 0.10$\pm$0.03 & 0.15\farcs{} &  \phn{}\phn{}$-$0.63$\pm$0.32 	&  					& iii. 		\\		
ZTF21abckwhn    &    \textit{SLSN?}					&    	 0.30$\pm$0.03 & 3.59\farcs{} &  $\textless-$0.04 	&  					& iii. 		\\		
ZTF21abfxibf    &    Ia								&    	 0.06$\pm$0.02 & 0.02\farcs{} &  $\textless-$0.71 	&  					& iv. 		\\
ZTF21abptxfk    &    II								&    	 0.10$\pm$0.03 & 3.09\farcs{} &  $\textless-$0.87 	&  					& iii. 		\\
ZTF21acfabut    &    Ia								&    	 0.07$\pm$0.02 & 2.86\farcs{} &  $\textless-$0.84 	&  					& i. 		\\
ZTF21achknpr    &    \textit{Ia}					&    	 0.29$\pm$0.03 & 2.99\farcs{} &  $\textless$~\phs{}0.11 	&  					& v. 		\\
\hline   
\hline
\multicolumn{7}{p{10cm}}{$^{a}$ Classifications given in italics are based on photometric templates or cross-matching. All other classifications are based on spectroscopic observations.} \\\multicolumn{7}{l}{$^{b}$ Colours are given corrected for Milky Way extinction only.} \tabularnewline
\end{tabular}
\end{center}
\end{table*}

\section{Intrinsically luminous assuming host redshift candidates}
\label{sect:appdx:bright_outliers}

Here we provide information on the candidates selected based on the inferred bright absolute magnitudes assuming the photometric redshift of the likely PS1 host galaxy.

\begin{table*}
\begin{center}
\caption{Final candidate glSNe selected by the luminosity-based method.}
\label{tab:bright_outliers}
\begin{tabular}{llccccccc}
\hline
\textbf{ZTF name}  	&	\textbf{Classification} 	& \textbf{Redshift}$^{a}$ & \textbf{Brightest $r$-mag}$^{b}$  &            & \multicolumn{2}{c}{\textbf{PS1 host galaxy}}   &      & \textbf{SALT2 $z$}$^{c}$ \\
                    &	                                &                   &                             & \textbf{photo-$z$} & \textbf{separation}      & \textbf{$i_{\textrm{Kron}}$} & \textbf{$M_r$}  &    \\
\hline
\hline
ZTF18acakmul &          &           & 19.47$\pm$0.27 & 0.573$\pm$0.063 & 3.79\farcs{} & 20.38 & $\leq$$-$22.15 & -                  \\
ZTF19acsvqrx & Ia       & 0.085     & 18.90$\pm$0.11 & 0.573$\pm$0.052 & 4.09\farcs{} & 20.42 & $\leq$$-$23.01 & 0.102$\pm$0.063    \\
ZTF20aaiftbn &          &           & 19.55$\pm$0.19 & 0.775$\pm$0.076 & 0.08\farcs{} & 21.75 & $\leq$$-$22.93 & -                  \\
ZTF20aauqecq &          &           & 19.31$\pm$0.11 & 0.612$\pm$0.069 & 1.76\farcs{} & 20.20 & $\leq$$-$22.54 & 0.137$\pm$0.038    \\
ZTF20aavcslg &          &           & 19.75$\pm$0.16 & 0.502$\pm$0.046 & 4.80\farcs{} & 19.96 & $\leq$$-$21.72 & 0.129$\pm$0.069    \\
ZTF20abewxcw &          &           & 19.33$\pm$0.13 & 0.406$\pm$0.045 & 3.79\farcs{} & 18.75 & $\leq$$-$21.52 & 0.128$\pm$0.048    \\
ZTF20abgbiet & Ia       & 0.060     & 17.94$\pm$0.05 & 0.499$\pm$0.032 & 1.40\farcs{} & 19.26 & $\leq$$-$23.98 & 0.027$\pm$0.015    \\
ZTF20abrnwfc &          &           & 19.72$\pm$0.12 & 0.607$\pm$0.059 & 0.14\farcs{} & 20.68 & $\leq$$-$22.40 & -                  \\
ZTF20acpzdof &          &           & 19.48$\pm$0.13 & 0.522$\pm$0.059 & 1.91\farcs{} & 20.74 & $\leq$$-$22.02 & 0.060$\pm$0.10     \\
ZTF20acwytxn & TDE      & 0.353     & 18.45$\pm$0.13 & 0.420$\pm$0.052 & 0.12\farcs{} & 19.24 & $\leq$$-$22.35 & -                  \\
ZTF21aabgjcz &          &           & 19.51$\pm$0.30 & 0.500$\pm$0.038 & 0.10\farcs{} & 19.71 & $\leq$$-$21.70 & -                  \\
ZTF21abcksdj &          &           & 19.37$\pm$0.10 & 0.605$\pm$0.054 & 2.84\farcs{} & 20.16 & $\leq$$-$22.70 & 0.150$\pm$0.110    \\
ZTF21abfqvjb & Ia       & 0.047     & 18.74$\pm$0.08 & 0.366$\pm$0.038 & 2.73\farcs{} & 19.37 & $\leq$$-$22.61 & 0.114$\pm$0.026    \\
ZTF21abfxhxm &          &           & 20.00$\pm$0.20 & 0.637$\pm$0.063 & 1.62\farcs{} & 20.53 & $\leq$$-$22.00 & -                  \\
ZTF21abgxbvi &          &           & 17.90$\pm$0.05 & 0.393$\pm$0.055 & 3.25\farcs{} & 19.07 & $\leq$$-$23.53 & -                  \\
ZTF21abhfzzs & Ia-91bg  & 0.031     & 18.24$\pm$0.13 & 0.314$\pm$0.061 & 1.52\farcs{} & 20.44 & $\leq$$-$22.74 & 0.026$\pm$0.038    \\
ZTF21abpiqsc &          &           & 18.85$\pm$0.06 & 0.684$\pm$0.074 & 4.34\farcs{} & 21.36 & $\leq$$-$24.03 & -                  \\
ZTF21abvpudz &          &           & 19.61$\pm$0.17 & 0.460$\pm$0.057 & 0.19\farcs{} & 19.55 & $\leq$$-$21.32 & -                  \\
ZTF21abvqvvf &          &           & 19.68$\pm$0.14 & 0.519$\pm$0.058 & 4.78\farcs{} & 19.29 & $\leq$$-$21.70 & 0.041$\pm$0.054    \\
\hline   
\hline
\multicolumn{9}{p{10cm}}{$^{a}$ Redshift of the classification spectrum reported on the TNS.} \\
\multicolumn{9}{l}{$^{b}$ Observed apparent magnitude of the brightest $r$-band detection in the light curve.} \tabularnewline
\multicolumn{9}{l}{$^{c}$ Redshift assuming a SALT2 template fit with $x_1 = c = 0$.} \tabularnewline
\end{tabular}
\end{center}
\end{table*}



\bsp	
\label{lastpage}
\end{document}